\pdfoutput=1

\documentclass[twocolumn]{aastex631}
\usepackage{epstopdf}
\usepackage{xspace}
\usepackage{graphicx}
\usepackage{amssymb}
\usepackage{amsmath}
\usepackage{natbib}
\usepackage{float}
\usepackage{hyperref}

\shorttitle{Magnetization of Current-Carrying Jets}
\shortauthors{Kr\'{o}l et al.}

\begin{document}

\title{Magnetization of Relativistic Current-Carrying Jets with Radial Velocity Shear}

\correspondingauthor{D.~{\L}.~Kr\'{o}l}
\email{dominika.l.krol@doctoral.uj.edu.pl}

\author[0000-0002-3626-5831]{Dominika~{\L}.~Kr\'{o}l}
\affiliation{Astronomical Observatory of the Jagiellonian University, Orla 171, 30-244 Krak\'{o}w, Poland}

\author[0000-0001-8294-9479]{{\L}ukasz~Stawarz}
\affiliation{Astronomical Observatory of the Jagiellonian University, Orla 171, 30-244 Krak\'{o}w, Poland}

\author[0000-0003-0936-8488]{Mitchell~C.~Begelman}
\affiliation{JILA, University of Colorado and NIST, 440 UCB, Boulder, CO 80309-0440, USA}
\affiliation{Department of Astrophysical and Planetary Sciences, 391 UCB, University of Colorado, Boulder, CO 80309-0391, USA}
\author{Jos\'{e}-Mar\'{\i}a Mart\'{\i}}
\affiliation{Departament d’Astronomia i Astrof\'{\i}sica, Universitat de Val\`{e}ncia, C/ Dr. Moliner, 50, 46100, Burjassot, Val\`{e}ncia, Spain}
\affiliation{Observatori Astron\`{o}mic, Universitat de Val\`{e}ncia, C/ Catedr\`{a}tic Jos\'{e} Beltr\'{a}n 2, 46980, Paterna, Val\`{e}ncia, Spain}
\author[0000-0003-2784-0379]{Manel Perucho}
\affiliation{Departament d’Astronomia i Astrof\'{\i}sica, Universitat de Val\`{e}ncia, C/ Dr. Moliner, 50, 46100, Burjassot, Val\`{e}ncia, Spain}
\affiliation{Observatori Astron\`{o}mic, Universitat de Val\`{e}ncia, C/ Catedr\`{a}tic Jos\'{e} Beltr\'{a}n 2, 46980, Paterna, Val\`{e}ncia, Spain}
\author{Bohdan~A.~Petrenko}
\affiliation{Taras Shevchenko National University of Kyiv, Volodymyrska St 60, 01033, Kyiv, Ukraine}

\begin{abstract}
Astrophysical jets, launched from the immediate vicinity of accreting black holes, carry away large amounts of power in a form of bulk kinetic energy of jet particles and electromagnetic flux. Here we consider a simple analytical model for relativistic jets at larger distances from their launching sites, assuming a cylindrical axisymmetric geometry with a radial velocity shear, and purely toroidal magnetic field. We argue that, as long as the jet plasma is in magnetohydrostatic equilibrium, such outflows tend to be particle dominated, i.e. the ratio of the electromagnetic to particle energy flux, integrated over the jet cross-sectional area, is typically below unity, $\sigma < 1$. At the same time, for particular magnetic and radial velocity profiles, magnetic pressure may still dominate over particle pressure for certain ranges of the jet radius, i.e. the local jet plasma parameter $\beta_{pl} < 1$, and this may be relevant in the context of particle acceleration and production of high-energy emission in such systems. The jet magnetization parameter can be elevated up to the modest values $\sigma \lesssim \mathcal{O}(10)$ only in the case of extreme gradients or discontinuities in the gaseous pressure, and a significantly suppressed velocity shear. Such configurations, which consist of a narrow, unmagnetized jet spine surrounded by an extended, force-free layer, may require an additional poloidal field component to stabilize them against current-driven oscillations, but even this will not elevate substantially their $\sigma$ parameter.
\end{abstract}

\section{Introduction}
\label{intro}

Relativistic jets found in various types of astrophysical sources of high-energy radiation, such as active galactic nuclei (AGN), microquasars, and gamma-ray  bursts, are believed to be formed via the efficient extraction of energy and angular momentum, in the form of Poynting flux, from the rotating black hole/accretion disk system (\citealt{Blandford77}; for an updated compendium see \citealt{Meier12}, also \citealt{Komissarov21} for a summary of recent developments in numerical simulations of jet launching). At the initial stages of their evolution, such structures are magnetically dominated, rapidly expanding, and only mildly relativistic. Thereafter collimation and acceleration start to proceed in accord, gradually converting the outflow to a fully-formed plasma jet, at the expense of the magnetic energy. However, in the framework of the ideal magnetohydrodynamical (MHD) description, and in the relativistic regime, such a conversion cannot be efficient, in the sense that the collimation and acceleration of the flow due to the magnetic tension and magnetic pressure gradient, respectively, are limited by the increasing electric force, which counter-balances the magnetic force. Efficient collimation and acceleration may, however, be reinforced by the external pressure, for example related to the star's interior in the case of gamma-ray bursts, or accretion disk winds in the case of AGN \citep[e.g.,][and references therein]{Lyubarsky09}.

Considering a power-law external pressure profile, \citet{Lyubarsky10} showed in particular that, for various initial magnetic field configurations or external pressure profiles, jets could possibly cease to be Poynting-flux dominated only at ``logarithmically large distances''. It is worth pointing out that the author defines a ``matter-dominated'' jet through the condition $\sigma < 0.1$, where $\sigma$ is the ratio of the jet magnetic energy flux to the particles' kinetic energy flux; the reason for this is that, in the intermediate regime $0.1< \sigma < 1$, formation of strong shocks within the outflow --- and hence energy dissipation as well as bulk deceleration of the flow due to interactions with the ambient medium --- proceed rather differently than in the purely hydrodynamical case \citep[e.g.,][]{Komissarov99,Kirk00}.

A contradictory conclusion follows from modelling of the observational data for relativistic jets in AGN, which often point towards very weak magnetization of the outflows, even at relatively close distances from the launching site. For example, various approaches to the spectral fitting of broad-band blazar emission typically give the estimate $\sigma \ll 1$ for distances $\lesssim 10^4 \, r_g$, where $r_g \simeq 10^{14} \, M_9$\,cm is the gravitational radius corresponding to the black hole mass $M_9 \equiv M_{\bullet} / 10^9 \, M_{\odot}$ (e.g., \citealt{Sikora09,Ghisellini10,Rueda14,Saito15}; but see also in this context \citealt{Sobacchi19}). This  would  be consistent  with  the  maximum  efficiency  of Poynting-to-matter energy flux conversion achieved relatively close to the jet base, followed by a steady and basically dissipation-free propagation of a relativistic matter-dominated outflow. Beyond the ideal MHD approximation, such a maximum efficiency could possibly be achieved with a help of dissipation processes, in particular magnetic reconnection, or the non-linear development of various MHD instabilities, rapidly converting the jet magnetic energy into kinetic energies of the jet particles \citep[e.g.,][]{Sikora05,Giannios06,Chatterjee19}.

Keeping in mind the results and findings summarized above, here we consider the simplest analytical model for a relativistic, current-carrying jet at a \emph{large distance} from the launching site, where the outflow can be considered as fully collimated and accelerated to terminal bulk velocities. In particular, we assume a cylindrical axisymmetric geometry for a non-rotating jet, with a radial velocity shear; moreover, we assume that the jet magnetic field has only a toroidal component, the jet is in \emph{magnetohydrostatic equilibrium}, and finally that the jet particles obey an ultra-relativistic equation of state.

In the framework of this approximation, introduced in more detail in Section\,\ref{sec:model}, we argue that the outflows tend to be matter-dominated, in the sense that the ratio of the jet Poynting and particle energy fluxes, integrated over the jet cross sectional area, i.e. the jet magnetization $\sigma$ parameter, is less than unity, as long as the radial pressure profiles are smooth with no extreme gradients or discontinuities. In Section\,\ref{sec:proof} we provide a simple analytical proof for this statement, valid however for only a certain class of magnetic field profiles, namely those for which the rest-frame magnetic pressure attains its global maximum at the jet boundary. In Section\,\ref{sec:numeric}, we explore further numerical solutions corresponding to various parametrizations of the generalized magnetic and velocity profiles, as well as various boundary conditions. There we observe that, even in the case of matter-dominated outflows with $\sigma < 1$, it is possible for the magnetic pressure to exceed particle pressure at certain ranges of the jet radii, or in other words for the local plasma beta parameter to alternate between $\beta_{pl} > 1$ and $\beta_{pl} < 1$ depending on the distance from the jet axis. We also explore cases with tangential discontinuities in the jet gaseous pressure, arguing that, for such, one may formally obtain magnetic-dominated outflows with $\sigma \lesssim \mathcal{O}(10)$, however at the price of huge pressure jumps by orders of magnitude. In the final Section\,\ref{sec:discussion}, we discuss our findings in the general context of particle acceleration in astrophysical jets, commenting also on the stability of the analyzed structures against current-driven oscillations, and on a role of an additional component of a poloidal magnetic field, in particular when confined to the narrow spine of the jet.

\section{The Jet Model}
\label{sec:model}

Let us consider a non-rotating, cylindrical axisymmetric jet in magnetohydrostatic equilibrium, for which the co-moving magnetic field $\vec{B'}$ is decomposed into the poloidal and toroidal components,
\begin{equation}
\vec{B'}_P = B'_r \, \hat{r} + B'_z \, \hat{z} \quad {\rm and} \quad \vec{B'}_T = B'_{\phi} \, \hat{\phi} \, .
\end{equation}
In this paper, we analyse a purely toroidal configuration, $B'_r = B'_z = 0$, due to the fact that, at large distances from the jet base, any poloidal magnetic field is typically expected to be negligible \citep[see][]{Begelman84}. The current $\vec{I}$ associated with such a jet is therefore parallel or anti-parallel to the jet axis, depending on the helicity of the jet magnetic field. Note that this would not be valid for a conical jet, for which there could also be a radial current component. However, at large distances from the jet base, effectively beyond the host galaxy, the background pressure should not be that of the interstellar medium, but instead of the hot, over-pressured jet cocoon, and as such should be independent of $z$, justifying our model assumption regarding a cylindrical jet geometry \citep[see in this context][]{Begelman89}. The only non-zero component of the current density, $\vec{J'} = J' \, \hat{z}$, is then
\begin{equation}
J' = {c \over 4 \pi r'} \,\, \partial_{r'}\!\left(r' B'_{\phi}\right) \, ,
\end{equation}
so that the magnetohydrostatic equilibrium condition $c \, \partial_{r'} P = - J' \, B'_{\phi}$, where $P$ denotes the comoving particle pressure, gives the radial profile
\begin{equation}
\partial_r P = - {1 \over 8 \pi r^2} \,\,\, \partial_r\!\!\left({r^2 B_{\phi}^2 \over \Gamma^2}\right) \, ,
\label{eq}
\end{equation}
where $B_{\phi} = B'_{\phi} \, \Gamma$ and $\Gamma = (1 - \beta^2)^{-1/2}$ is the bulk Lorentz factor of the portion of the fluid with the velocity $\beta$ \citep[see, e.g.,][]{Appl92}. The jet current, $I = \frac{1}{2} \, c \, r \, B_{\phi}\!(R_j)$, where $R_j$ is the jet radius, has to be compensated by a return current $I_{\rm ret}$ at larger scales where the outflow terminates.

Let us denote the proper enthalpy and the proper internal energy density of the jet particles by $w$ and $u$, respectively. In the case of ultra-relativistic jet particles, i.e. for $P = u /3$, one has $w = P + u = 4 P$, the case we analyze exclusively hereafter. With such, the energy flux associated with the jet particles is
\begin{equation}
L_p = 2 \pi \, c \, \int\!dr \, r \, \Gamma^2 \beta \, w = 8 \pi \, c \, \int\!dr \, r \, \beta \, \Gamma^2 \, P \, ,
\end{equation}
while the jet Poynting flux is
\begin{equation}
L_B = 2 \pi \, c \, \int\!dr \, r \, {E_r B_{\phi} \over 4 \pi} = {1 \over 2} \, c \, \int\!dr \, r \, \beta  \, B_{\phi}^2  \, ,
\end{equation}
where $E_r = \beta B_{\phi}$ is the radial component of the jet electric field. The jet magnetization parameter $\sigma$ is then
\begin{equation}
\sigma \equiv {L_B \over L_p} = {1 \over 16 \pi} \, {\int\!dr \, r \, \beta \, B_{\phi}^2 \over \int\!dr \, r \, \beta \, \Gamma^2 \, P} \, .
\label{sigma_def1}
\end{equation}

It is convenient to introduce the dimensionless parameter $x \equiv r / R_j$, and the normalized profiles
\begin{eqnarray}
b\!(x)  \equiv \frac{B_{\phi}\!(x)}{B_{\phi}\!(1)} & \, : \quad & b\!(0)=0 \, , \quad b\!(1)\equiv 1 \, , \quad {\rm and} \\
p\!(x)  \equiv \frac{P\!(x)}{P\!(1)}  & \, : \quad & p\!(0) >0 \, , \quad p\!(1)\equiv 1 .
\label{bgp}
\end{eqnarray}
For now, we do not make any assumption regarding the particular monotonicity of $b\!(x)$ or $p\!(x)$ profiles, but only that the magnetic field has a particular helicity all along the outflow, and that it does not vanish at the jet boundary, $B_{\phi}(1)>0$; note however that, for a purely toroidal configuration, magnetic field always drops to zero at the jet axis, $B_{\phi}(0)=0$. As for the velocity profile, we consider only the cases with the jet bulk Lorentz factor $\Gamma\!(x)$ decreasing monotonically along the jet radius starting from $\Gamma\!(0)\equiv \Gamma_0>1$ down to $\Gamma(1)=1$, excluding anomalous shear layers \citep[see in this context][]{Aloy06,Mizuno08}.

Next we introduce the parameter
\begin{equation}
q \equiv \frac{B_{\phi}^2\!(1)/8\pi}{P\!(1)}\equiv\frac{P_B\!(1)}{P\!(1)}\equiv \beta^{-1}_{pl}\!(1) \, ,
\end{equation}
which is the ratio of the magnetic and gas pressures at the jet boundary, i.e. the inverse of the plasma beta parameter at $x=1$. To simplify the notation, we further introduce
\begin{equation}
f\!(x) \equiv \frac{b^2\!(x)}{\Gamma^2\!(x)} \, : \quad f\!(0)=0 \, , \quad f\!(1)=1 \, ,
\label{ff}
\end{equation}
which is basically the normalized rest-frame magnetic pressure, $f\!(x) \propto P_B(x) \equiv {B'_{\phi}}^2/8\pi$. 

With the above definitions, for a given set of profiles $b\!(x)$ and $\Gamma\!(x)$, or equivalently $f\!(x)$ and $\Gamma\!(x)$, and with a given boundary condition $q$, the magnetohydrostatic equilibrium condition, given in equation\,\ref{eq}, determines the particle pressure profile
\begin{equation}
    \label{pressure}
    p\!(x)=1+q-q\,f\!(x)+2q \int\limits_x^1\textrm{d}s\,\frac{f\!(s)}{s}
\end{equation}
Importantly, the requirement $\forall_{x\in[0,1]}: p\!(x) > 0$, translates to the constraint
\begin{equation}
\label{condition}
1+q^{-1}>f\!(x)-2\int\limits_x^1\textrm{d}s\,\frac{f\!(s)}{s}.
\end{equation}

\section{Jet $\sigma$ Parameter}
\label{sec:proof}

It was noted by \citet{Lisanti07} that, for the jet model setup as considered here, numerical solutions to the jet magnetization parameter always return $\sigma < 1$. Below we draft a simple analytical proof for this statement, valid however for only a certain class of magnetic profiles. In particular, in this section we consider only those magnetic profiles, for which the comoving magnetic field pressure is continuous and attains its maximum at the jet boundary, or in other words for which $f\!(x) < 1$ for every $x<1$, even though $f\!(x)$ is not necessarily monotonic within the entire range $x\in [0,1]$.

First, let us introduce the function
\begin{equation}
    h\!(x)\equiv x\,\beta\!(x)\,\Gamma^2\!(x) \, ,
\end{equation}
such that $h\!(0)= h\!(1)=0$, which has exactly one maximum at a given $x_0\in(0,1)$ for the assumed monotonically decreasing $\Gamma\!(x)$. With such, the jet $\sigma$ parameter can be re-written as:
\begin{equation}
\label{sigmaEq}
    \sigma=\frac{q}{2} \,\, \frac{\int\limits_0^1\textrm{d}x \,h(x)\,f(x)}{\int\limits_0^1\textrm{d}x \,h(x)\,p\!(x)} \, .
\end{equation}

We also define
\begin{equation}
    H\!(x)\equiv \frac{\int\limits_0^x\textrm{d}y \, h\!(y)}{\int\limits_0^1\textrm{d}z \, h\!(z)} \, ,
\end{equation}
which, by definition, is monotonically increasing from $H\!(0)=0$ to $H\!(1)=1$. Note that
\begin{equation}
    \frac{x\, H'\!(x)}{H\!(x)}\equiv \frac{\textrm{d}\ln H\!(x)}{\textrm{d}\ln x}=\frac{x\, h\!(x)}{\int\limits_0^x\textrm{d}y\,h\!(y)} \, .
\end{equation}
Moreover, since $\beta\!(x)\Gamma^2\!(x)$ is a continuous, monotonically decreasing function, from the Mean Value Theorem (MVT) we know that for a given fixed $x$, there is always $\xi\in[0,x]$ such that $\int\limits_0^x\textrm{d}y\,y \, \beta\!(y) \, \Gamma^2\!(y)=\beta\!(\xi)\, \Gamma^2\!(\xi) \, \int\limits_0^x\textrm{d}y \, y=\frac{1}{2}x^2\beta\!(\xi)\,\Gamma^2\!(\xi)$, and therefore
\begin{equation}
\label{Hfunc}
    \frac{x\,H'\!(x)}{H\!(x)}=\frac{x^2\beta\!(x)\,\Gamma^2\!(x)}{\frac{1}{2}x^2\beta\!(\xi)\,\Gamma^2\!(\xi)}\leq2 \, .
\end{equation}
In fact, $x \, H'\!(x)/H\!(x)$ is always monotonically decreasing from $2$ at $x=0$ down to $0$ at $x=1$. 

Using the form of the $\sigma$ parameter as given in equation\,\ref{sigmaEq}, along with the solution for the pressure profile \ref{pressure} and the condition for a positive gas pressure \ref{condition}, we obtain
\begin{equation}
    \label{ineq1}
    \frac{1}{2\sigma}>\frac{f(x)-2\int\limits_x^1\textrm{d}s\,\frac{f(s)}{s}-\langle f\rangle+\langle2\int\limits_y^1\textrm{d}s\,\frac{f(s)}{s}\rangle}{\langle f\rangle} \, ,
\end{equation}
where the averaging is over the $h(x)$ distribution, namely
\begin{equation}
    \langle X\rangle \equiv \int\limits_0^1\textrm{d}H \, X\equiv \frac{\int\limits_0^1\textrm{d}y\,h(y)\,X}{\int\limits_0^1\textrm{d}y\,h(y)} \, ,
\end{equation}
with $\int\textrm{dH}\equiv \int\textrm{d}x\,H'$. Hence if the right-hand side of equation (\ref{ineq1}) is larger than $1/2$ for every $x$, then $\sigma<1$.

In the case of $f(x)$ attaining its maximum at the jet boundary, as considered in this section, the condition (\ref{ineq1}) at $x=1$ reads as
\begin{equation}
    \label{ineq2}
    \frac{1}{2\sigma}>\frac{1-\langle f\rangle+\langle 2\int\limits_y^1\textrm{d}s\, \frac{f(s)}{s}\rangle}{\langle f\rangle} \, .
\end{equation}
Meanwhile, from the MVT we know that there is $\eta\in(0,1)$ such that $\langle f \rangle=f\!(\eta)<1$. Hence
\begin{eqnarray}
  &&  \biggl< 2\!\int\limits_y^1\!\textrm{d}s\,\frac{f(s)}{s}\biggr> = 2\!\int\limits_0^1\!\textrm{d}y \, H' \int\limits_y^1\!\textrm{d}s\,\frac{f(s)}{s}=2\!\int\limits_0^1\!\textrm{d}y\, H(y)\,\frac{f(y)}{y} \nonumber\\
    &&= 2\, \frac{H}{yH'}\bigg|_{y=\zeta}\int\limits_0^1\!\textrm{d}H \, f=2\,\frac{H}{yH'}\bigg|_{y=\zeta}\, \langle f\rangle\geq \langle f \rangle \, ,
\end{eqnarray}
where we used again the MVT with a certain $\zeta \in (0,1)$, and the fact that $H/(yH') \leq 1/2$ for every $y\in(0,1)$ (equation\,\ref{Hfunc}). All in all, we therefore have
\begin{equation}
    \frac{1}{2\sigma}>\frac{1}{f\!(\eta)}>1 \, ,
\end{equation}
resulting in $\sigma<\frac{1}{2}$.

\section{Jet Radial Profiles}
\label{sec:numeric}

In the framework of our simple model for current-carrying jets in magnetohydrostatic equilibrium, outlined in Section~\ref{sec:model}, the jet magnetization parameter $\sigma$ is always less than unity, as long as the comoving toroidal magnetic field pressure is continuous and attains its maximum at the jet boundary; a simple proof of this statement has been presented in the previous Section~\ref{sec:proof}. More complex cases, with pressure discontinuities and sharp maxima of the magnetic pressure located well within the jet body, can be investigated numerically,  by adopting various parametrizations of the $b(x)$ and $\Gamma(x)$ profiles.

\subsection{Continuous Profiles}

As noted in Section\,\ref{sec:model}, the anticipated magnetohydrostatic equilibrium condition determines the jet particle pressure profile, $p(x)$, for a given magnetic pressure and jet bulk Lorentz factor profiles $b(x)$ and $\Gamma(x)$, as well as a given boundary condition $q \equiv \beta_{pl}^{-1}\!(1)$. Let us therefore consider the three illustrative cases of certain $b(x)$ and $\Gamma(x)$ parametrizations, corresponding to the three different behaviours of the function $f(x)$, namely (a) the case with monotonically increasing $f(x)$, (b) the case with one pronounced global maximum of $f(x)$ at $x<1$, and (c) the case with multiple local maxima of $f(x)$ throughout the entire range of $x$.

\begin{figure}[!t]
    \centering
    \includegraphics[width=\columnwidth]{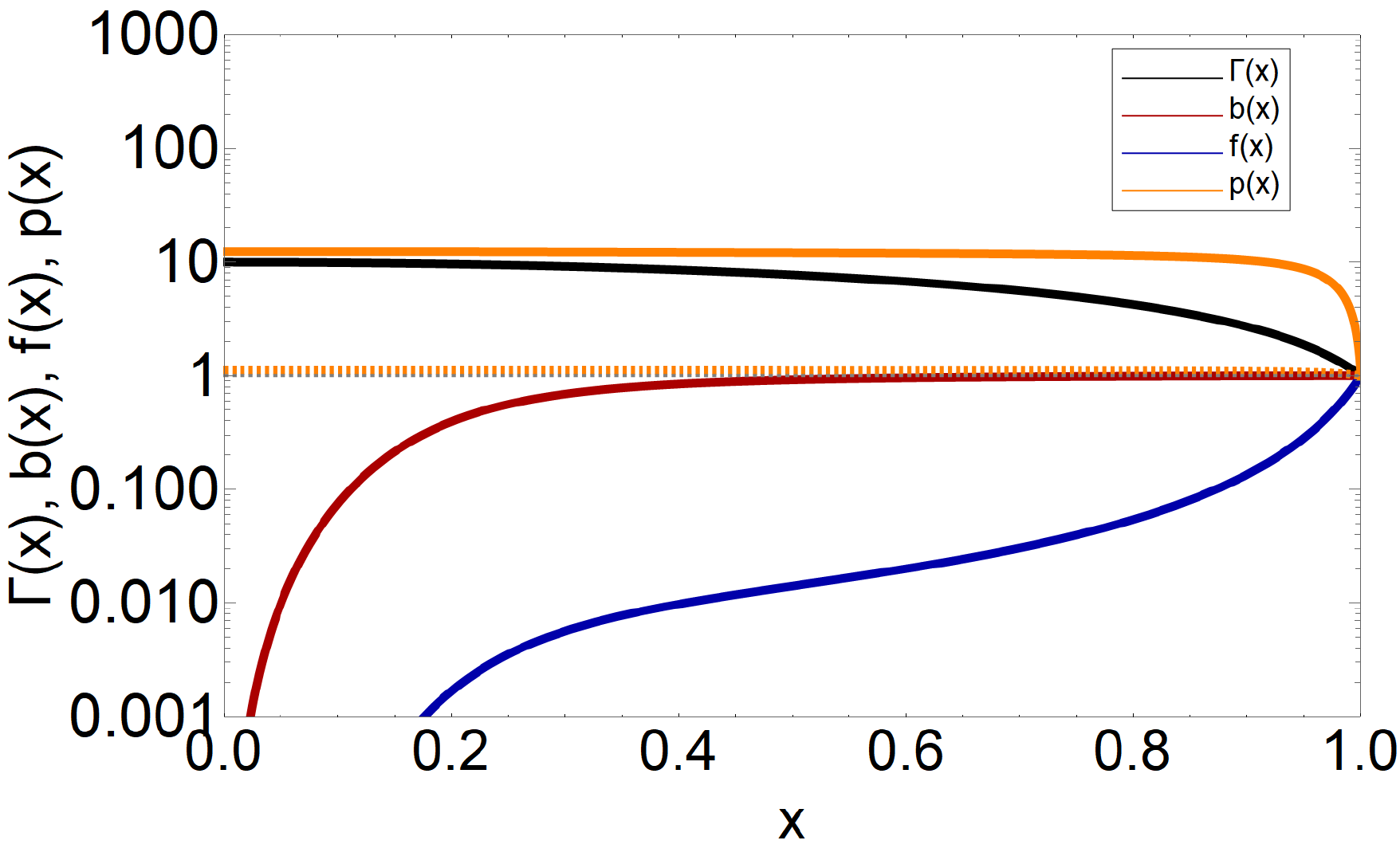}
     \includegraphics[width=\columnwidth]{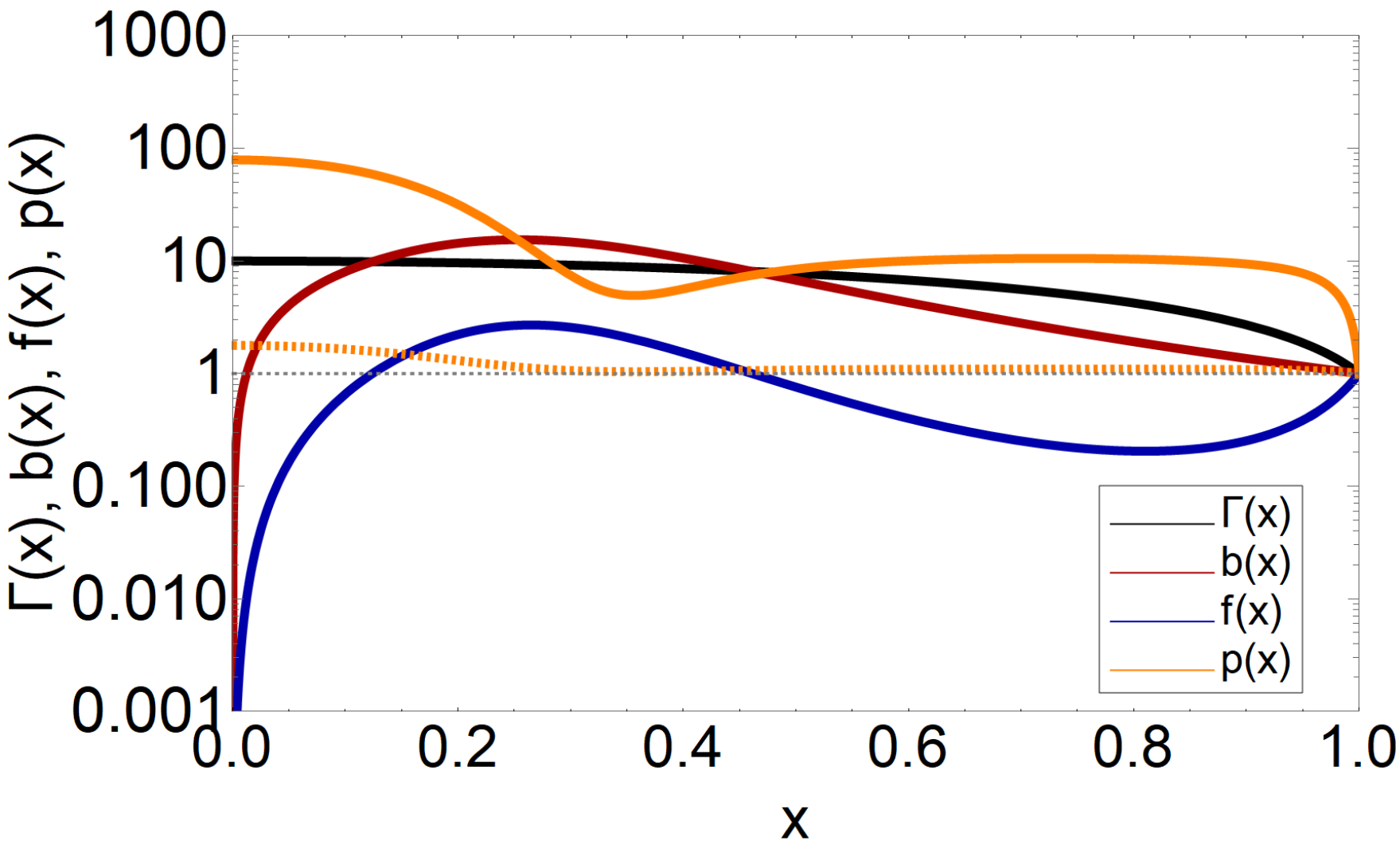}
    \includegraphics[width=\columnwidth]{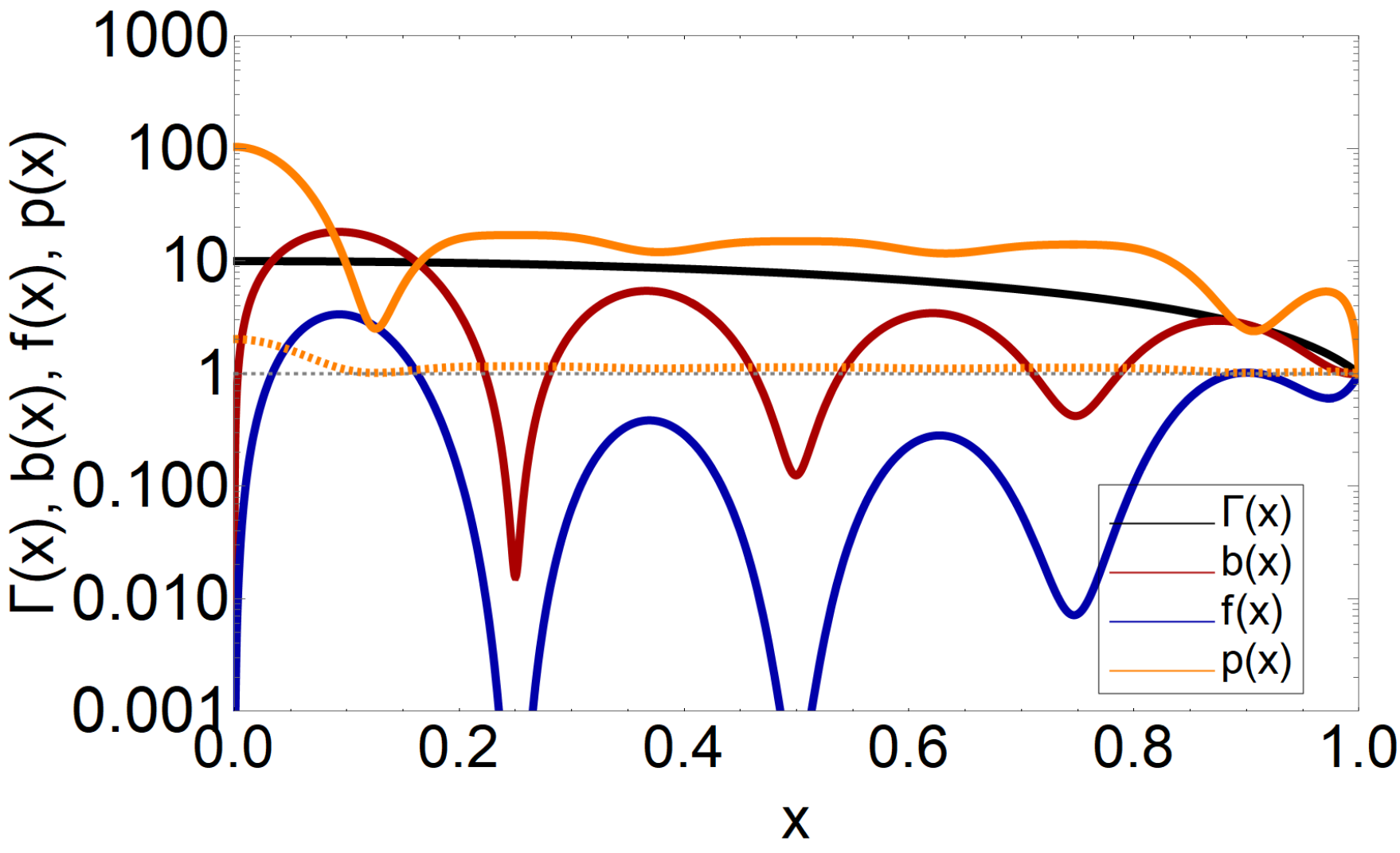}
    \caption{Three exemplary sets of $b(x)$ and $\Gamma(x)$ profiles, along with the corresponding $f(x)$ profiles, and the resulting pressure profiles $p(x)$ calculated for the boundary conditions $q=0.1$ (dashed curves) and $q=10$ (solid curves).}
    \label{fig:profiles}
\end{figure}

In the following numerical analysis, for clarity we follow one particularly simple paramerization of the jet bulk Lorentz, namely
\begin{equation}
\label{numerical_Gb_profiles}
\Gamma(x) = 1+(\Gamma_0-1)\times (1-x^{k}) \, ,
\end{equation}
with $\Gamma_0=10$ and $k=2$. For the magnetic profile $b(x)$, on the other hand, we consider various analytical prescriptions, selected to ensure positive particle pressure at all jet radii. In Figure\,\ref{fig:profiles} we present the three exemplary sets of $b(x)$ and $\Gamma(x)$ profiles, along with the corresponding $f(x)$ profiles and the resulting pressure profiles $p(x)$ calculated for $q=0.1$ and $q=10$.

As shown, the particle pressure is maximized at the jet axis, and, for monotonically increasing $f(x)$, it decreases monotonically toward the jet boundary. In the case of more complex magnetic pressure profiles, local maxima in the $b(x)$ profile correspond to local minima in $p(x)$. In general, sharp gradients in the comoving magnetic pressure with $f(x)$ locally exceeding unity, often generate exponential drops in particle pressure, leading to un-physical solutions with $p(x)<0$ for some ranges of the jet radius.

\begin{figure}[!t]
    \centering
    \includegraphics[width=\columnwidth]{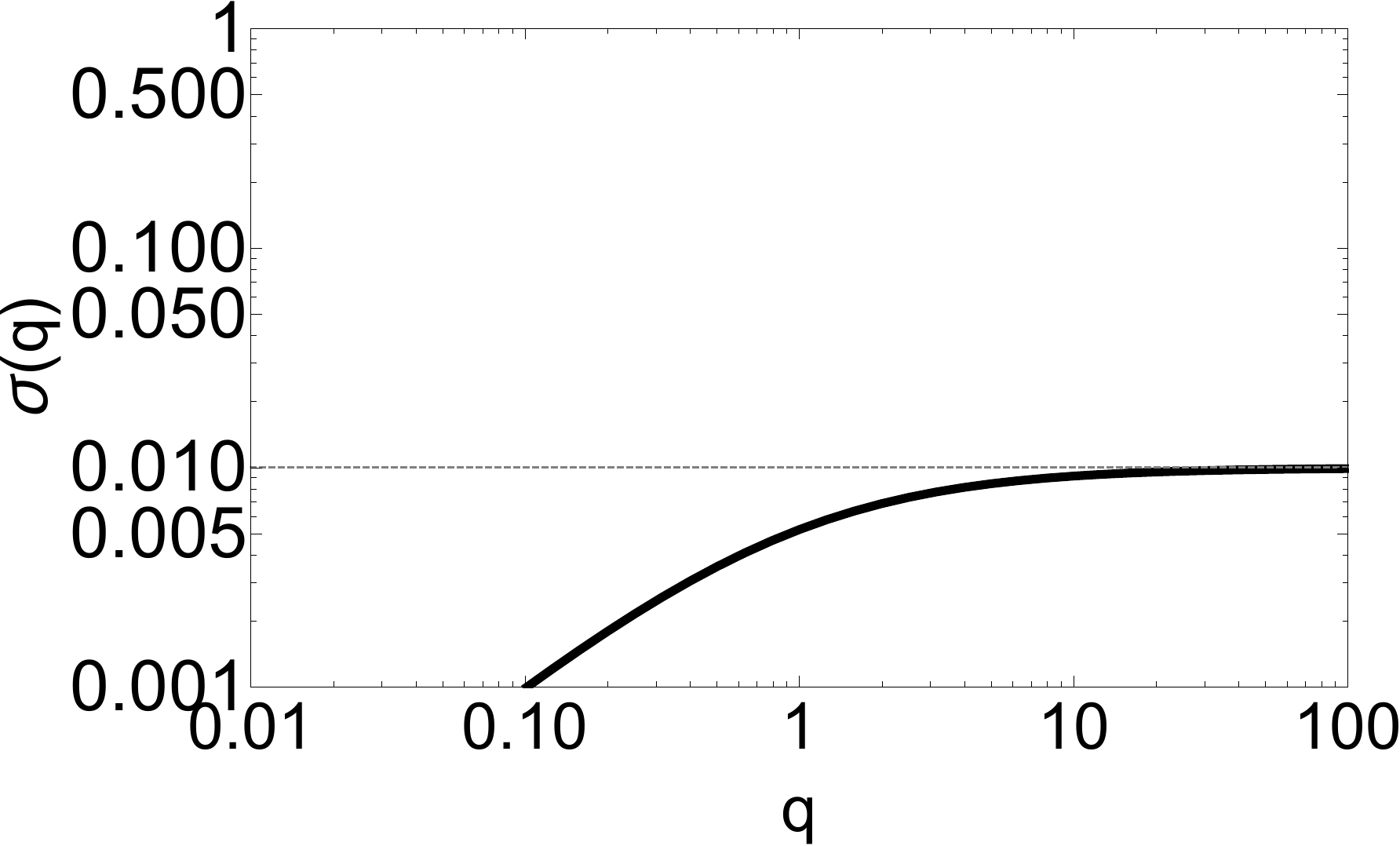}
     \includegraphics[width=\columnwidth]{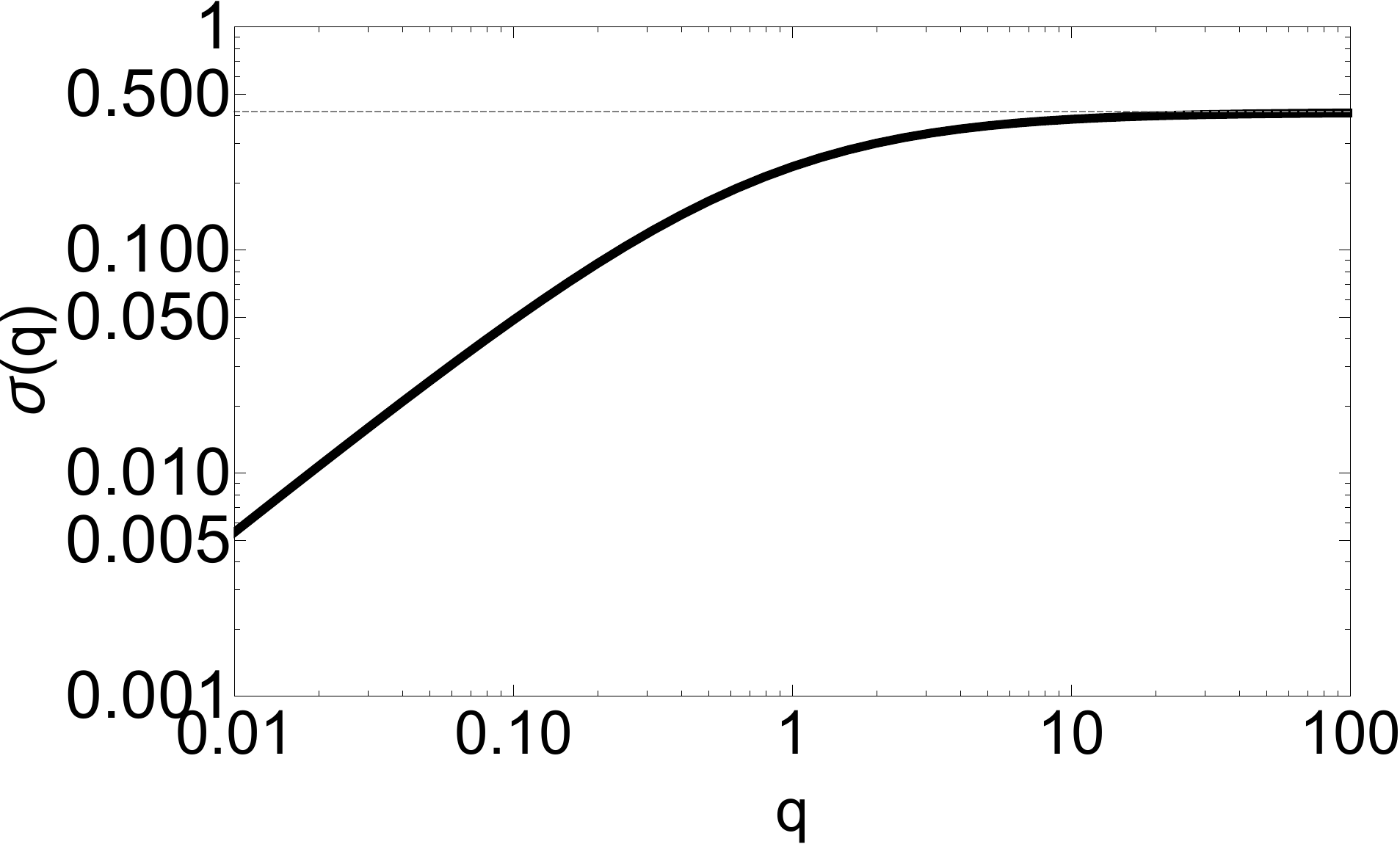}
    \includegraphics[width=\columnwidth]{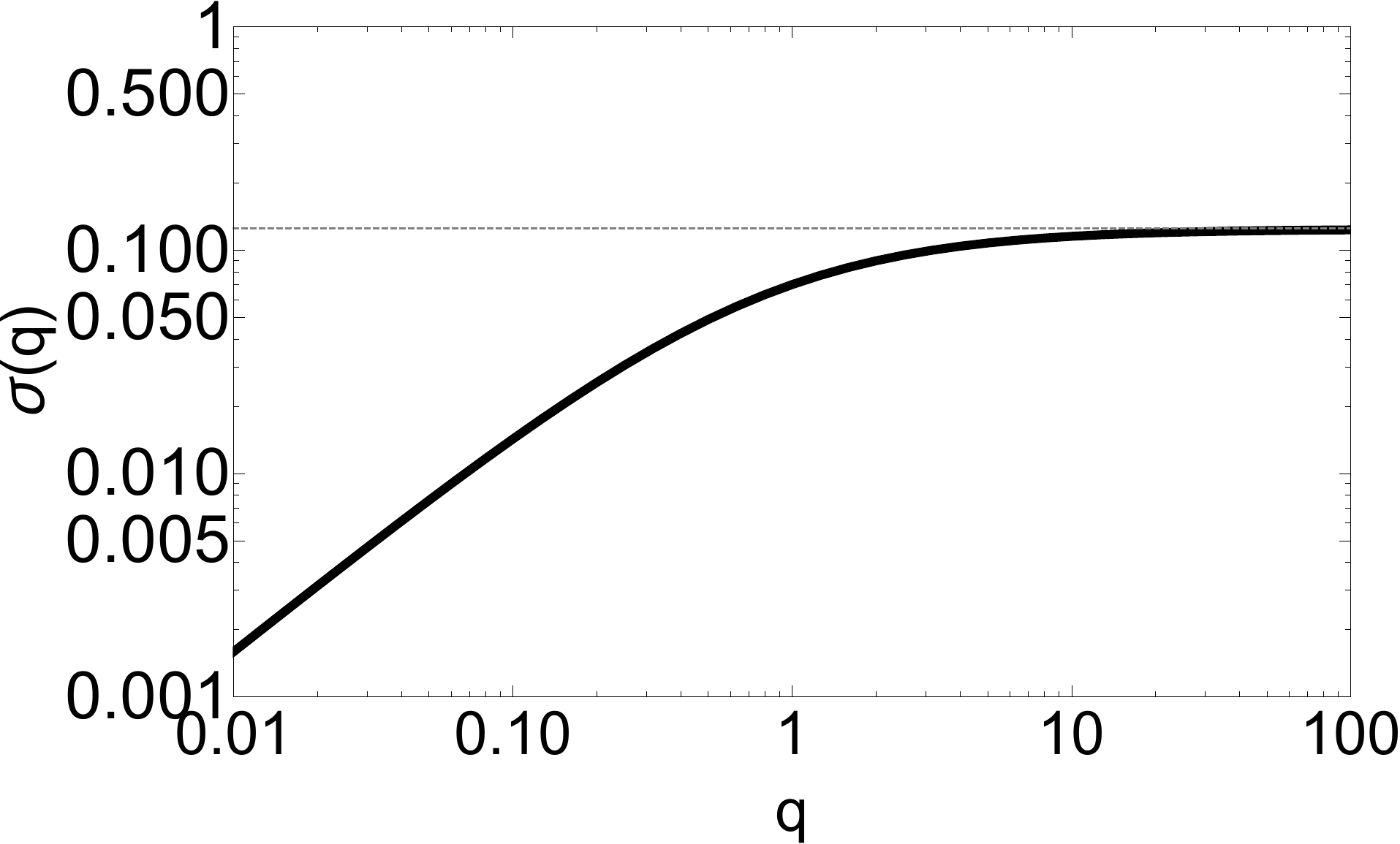}
    \caption{Jet magnetization parameter $\sigma$ calculated for different boundary condition values $q \equiv \beta_{pl}^{-1}\!(1)$; the three panels from top to bottom correspond to the three illustrative sets of $b(x)$ and $\Gamma(x)$ profiles shown in Figure\,\ref{fig:profiles}.}
    \label{sigma}
\end{figure}

Figure\,\ref{fig:profiles} also indicates how the boundary condition $q$ impacts the resulting jet particle pressure profiles, and hence the overall jet magnetization. Figure\,\ref{sigma} illustrates the latter effect explicitly, by providing the exact $\sigma$ values corresponding to different values of $q \in (0.01,100)$. As shown, starting from small $q$, i.e. negligible magnetic pressure at the jet boundary (with respect to the particle pressure), the jet magnetization $\sigma$ increases monotonically with increasing $q$, but only until $q\gtrsim 1$; from that point, the $\sigma$ parameter saturates and remains effectively constant (always below the unity), regardless of the ever-increasing $q> 10$.

\begin{figure}[!th]
    \centering
    \includegraphics[width=\columnwidth]{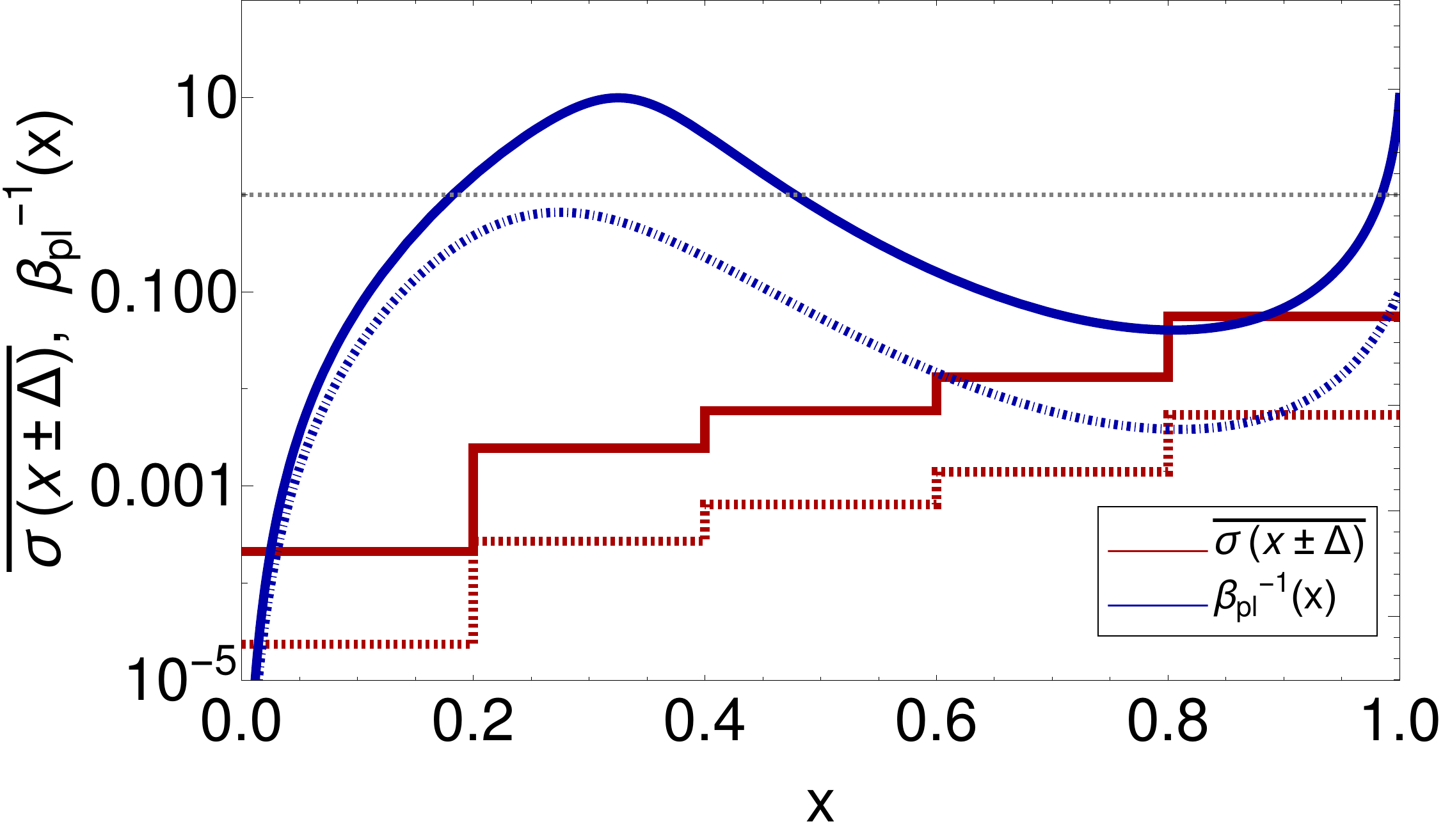}
     \includegraphics[width=\columnwidth]{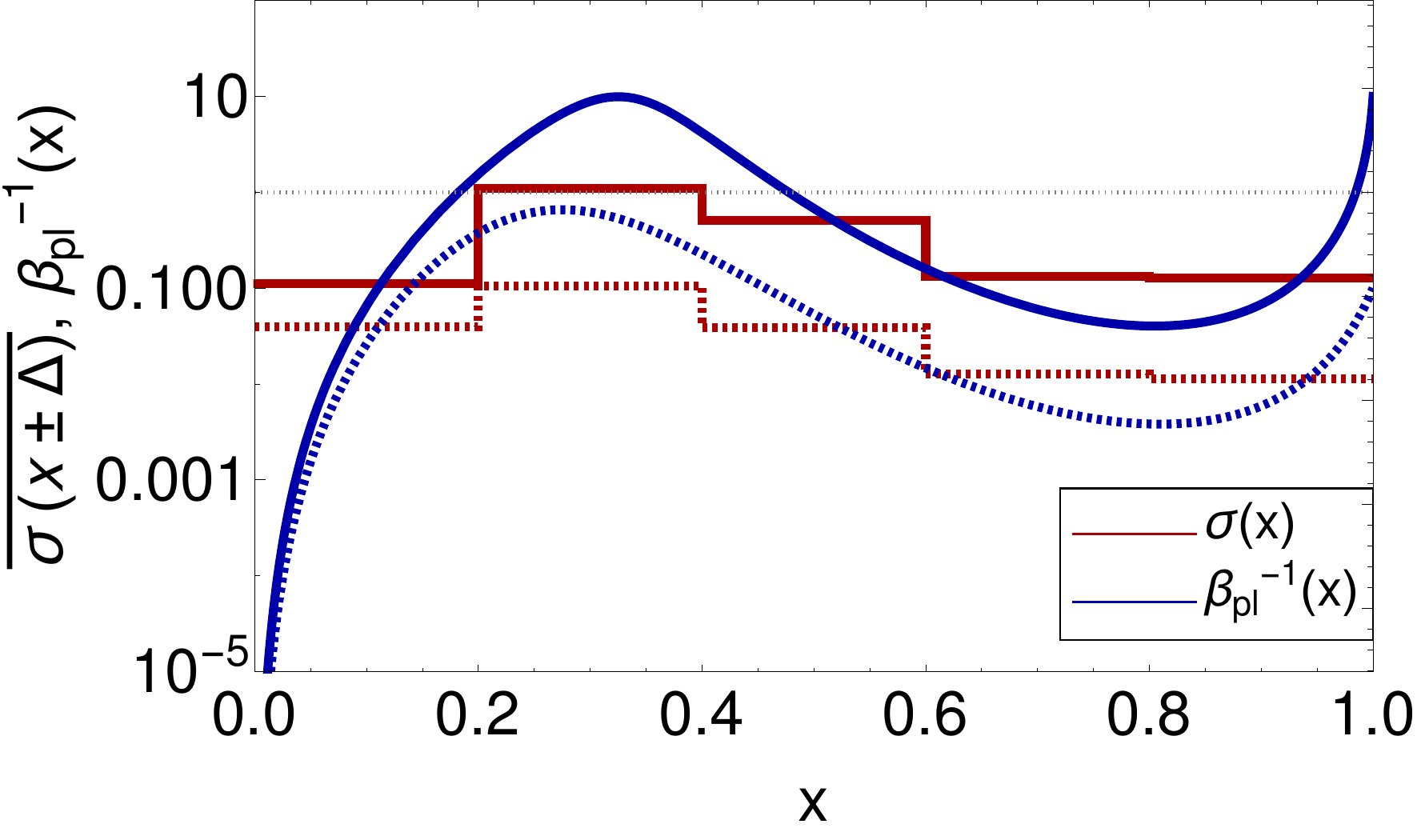}
    \includegraphics[width=\columnwidth]{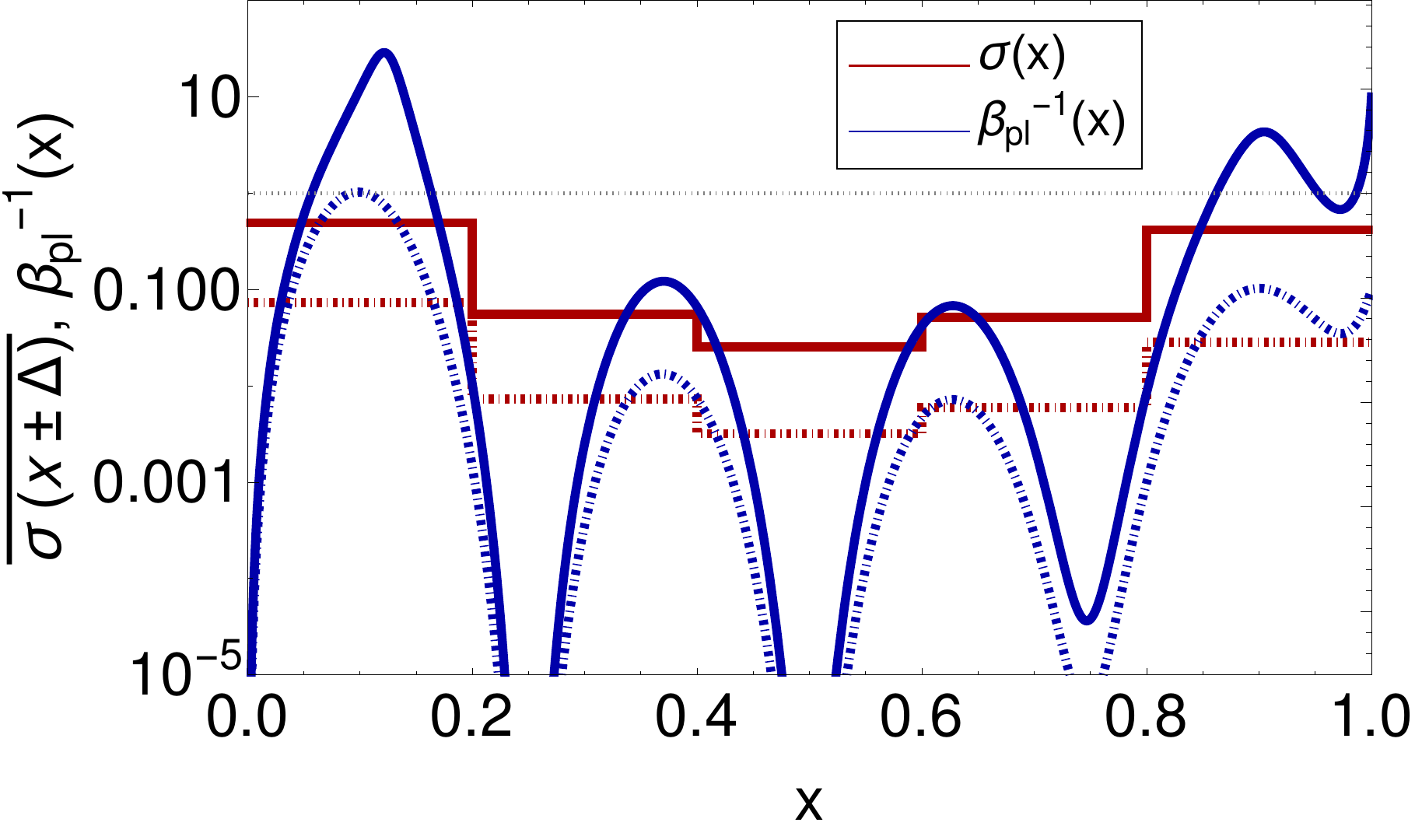}
   \caption{Jet plasma magnetization parameters $\beta_{pl}^{-1}(x)$ and $\overline{\sigma(x\pm\Delta)}$ with midpoints of averaging $x=\{0.1,0.3,0.5,0.7,0.9\}$ and $\Delta=0.1$, calculated for the boundary condition values $q = 0.1$ (dashed lines) and $q=10$ (solid lines); the three panels from top to bottom correspond to the three illustrative sets of $b(x)$ and $\Gamma(x)$ profiles shown in Figure\,\ref{fig:profiles}.}
    \label{beta_sigL}
\end{figure}

\begin{figure}[!t]
	\centering
	\includegraphics[width=\columnwidth]{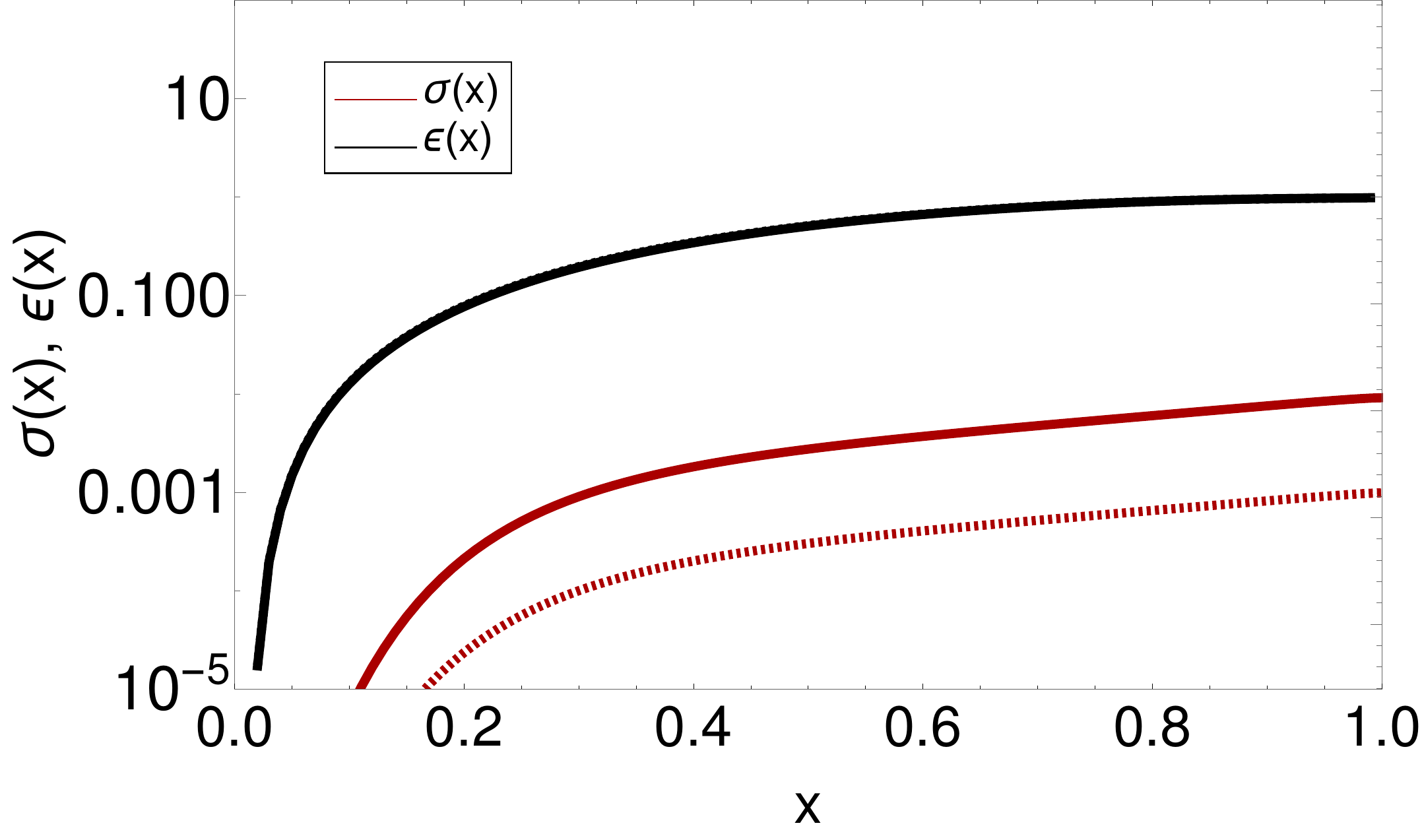}
	\includegraphics[width=\columnwidth]{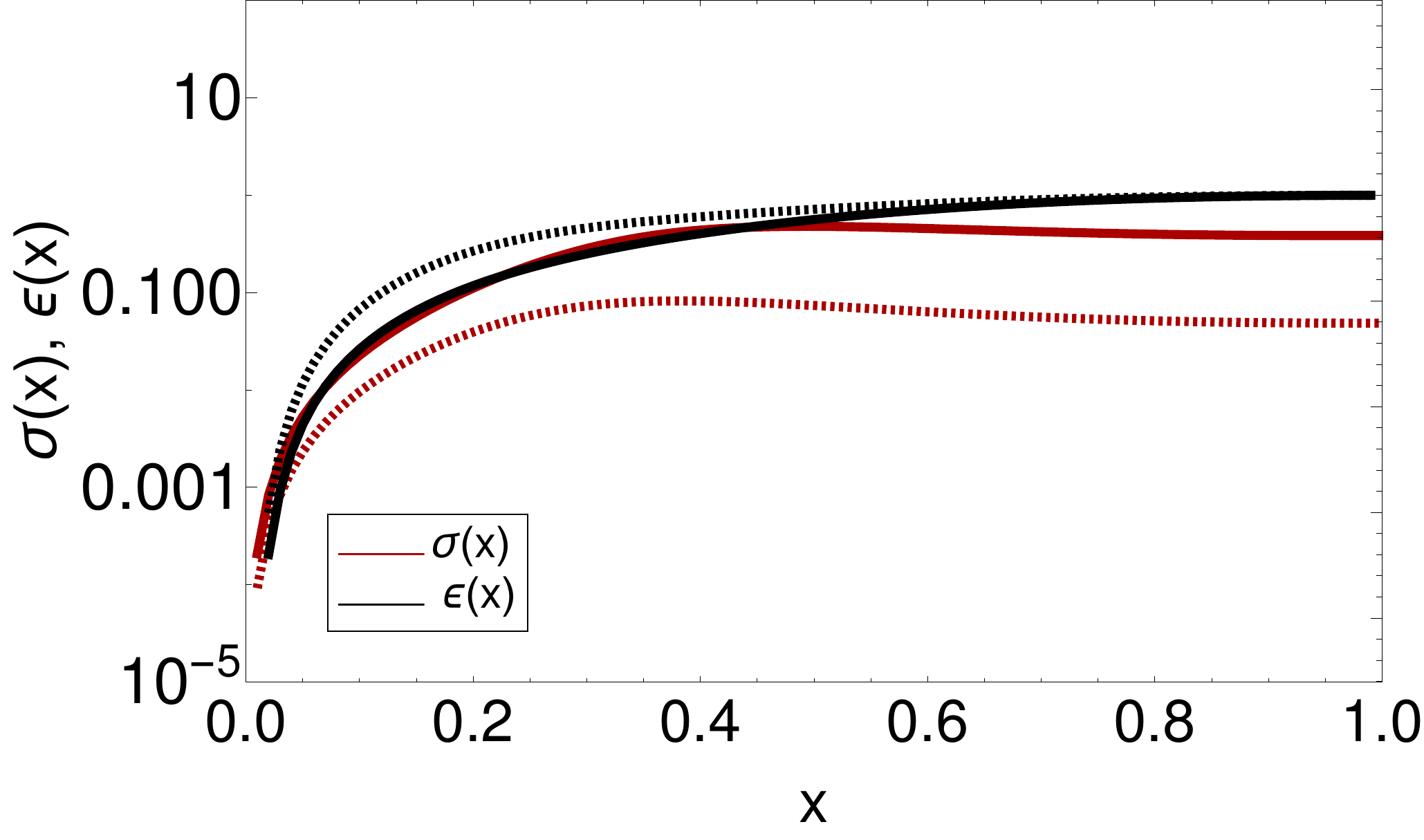}
	\includegraphics[width=\columnwidth]{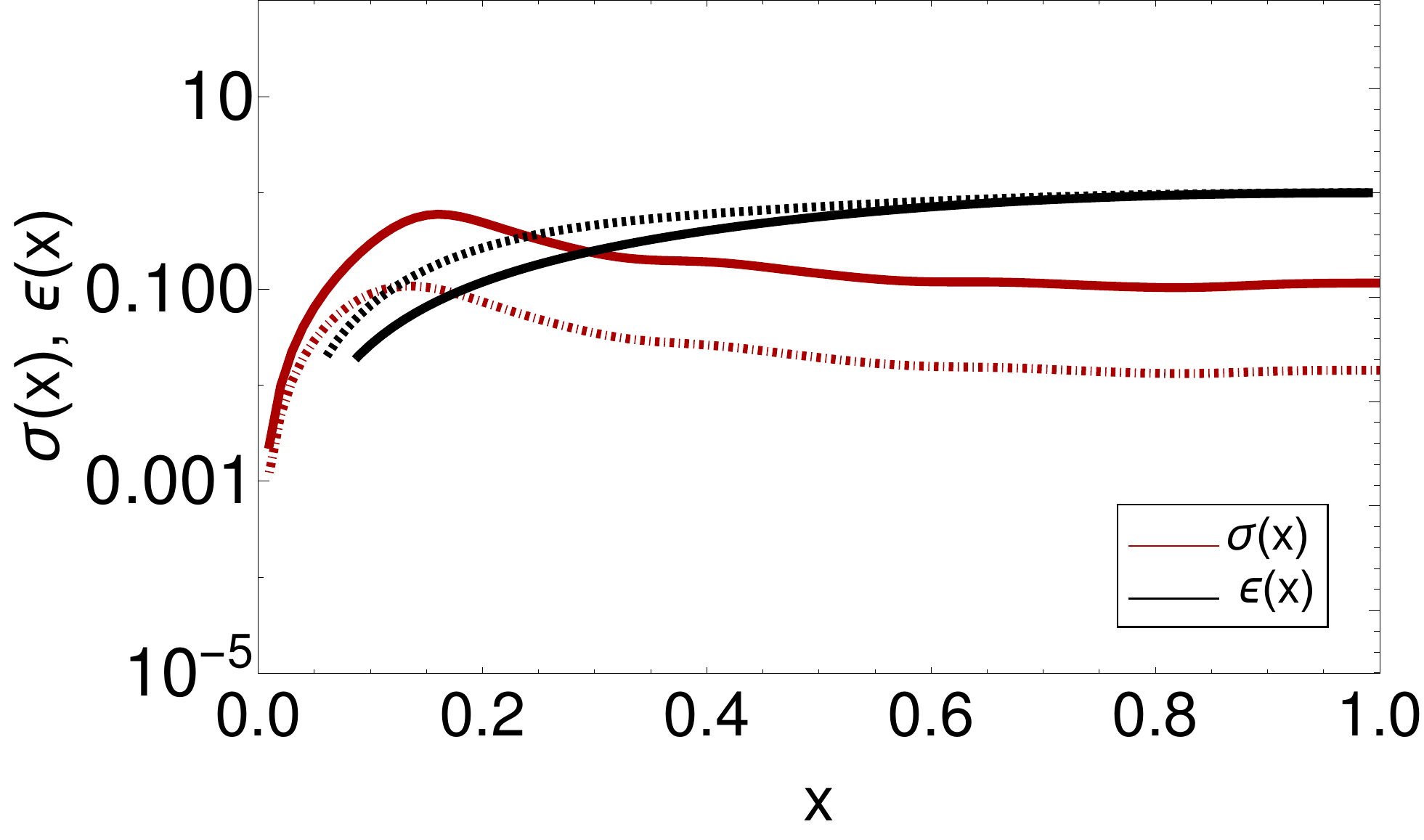}
	\caption{Cumulative $\sigma_c(x)$ parameter and the share of a total energy flux $\epsilon(x)$, calculated for the boundary condition values $q = 0.1$ (dashed lines) and $q=10$ (solid lines); the three panels from top to bottom correspond to the three illustrative sets of $b(x)$ and $\Gamma(x)$ profiles shown in Figure\,\ref{fig:profiles}.}
	\label{beta_sigK}
\end{figure}

Magnetohydrostatic equilibrium, assumed here for the jet, implies that any spatial changes in the magnetic pressure across the jet must be counter-balanced by the changes in the particle pressure and the magnetic tension. As a result, the rest-frame plasma beta parameter 
\begin{equation}
    \beta_{pl}^{-1}\!(x) \equiv  \frac{P_B\!(x)}{P\!(x)} = \frac{q \, f\!(x)}{p\!(x)} \, , 
\end{equation}
may change quite dramatically along the jet radius. Similarly, the ``local'' jet magnetization $\sigma$ parameter calculated for different ``layers'' of a sheared outflow with thickness $\Delta$, namely
\begin{equation}
\label{sigmaX}
\overline{\sigma(x\pm \Delta)} \equiv
\frac{\langle f\!(x) \rangle_{\Delta}}{\langle p\!(x) \rangle_{\Delta}} \equiv \frac{q}{2} \,\, \frac{\int\limits_{x-\Delta}^{x+\Delta}\textrm{d}y \, h\!(y) \, f\!(y)}{\int\limits_{x-\Delta}^{x+\Delta}\textrm{d}y \, h\!(y) \,p\!(y)} \, ,
\end{equation}
may also vary across the jet. Those changes, corresponding to the three sets of the $b(x)$ and $\Gamma(x)$ profiles considered in this sub-section, are shown in Figure\,\ref{beta_sigL}, for the two selected boundary condition values $q=0.1$ and 10, with midpoints of averiging $x=\{0.1,0.3,0.5,0.7,0.9\}$ and $\Delta=0.1$.

In the case of a monotonic $f(x)$, both $\beta_{pl}^{-1}(x)$ and $\sigma(x)$ increase monotonically from the jet spine toward the jet boundary. Models for which the comoving magnetic pressure attains a local maximum with $f(x) \gtrsim 1$ are, on the other hand, characterized by $\beta_{pl}^{-1}(x)$ exceeding unity at the approximately corresponding jet radii; the local $\sigma(x)$ parameter may then also approach unity for certain ranges of $x$, depending on the considered layer thickness $\Delta$. In other words, for such complex models with alternating currents, we observe the presence of extended domains within the outflow in which the comoving jet magnetic pressure dominates over the gaseous pressure, while the jet Poynting energy flux is comparable to the kinetic energy flux of the jet particles, even though the global $\sigma$ value (i.e., the value obtained after integrating over the entire cross-sectional area of the jet) is much below unity. 

Interestingly, one can give examples of particular magnetic pressure profiles with multiple maxima across the jet, for which the plasma beta parameter alternates between $>1$ and $<1$, resulting in a substantial jet radial stratification with respect to not only the bulk velocity of the jet plasma, but also the jet plasma magnetization. The question is, what relative amount of the total energy of the outflow is carried by the layers with such  vastly different magnetizations. 

In order to investigate this issue, for the same three illustrative magnetic and velocity profiles, and the boundary conditions $q=0.1$ and $q=10$, we also calculate the ``cumulative'' $\sigma_c(x)$, defined as
\begin{equation}
\label{sigmaC}
\sigma_c(x) \equiv \frac{q}{2} \, \frac{\int\limits_0^x\textrm{d}y \,  h\!(y) \, f\!(y)}{\int\limits_0^x\textrm{d}y \, h\!(y) \, p\!(y)} \, ,
\end{equation}
as well as 
\begin{equation}
    \epsilon(x) \equiv \frac{L_p(x)+L_B(x)}{L_p(1)+L_B(1)} \, ,
\end{equation}
which represents the fraction of the total energy flux integrated from $0$ up to a given $x$. Those are shown in Figure\,\ref{beta_sigK}. By comparing the $\sigma_c(x)$ and $\epsilon(x)$ profiles with the plasma beta parameter profiles we conclude that there potentially \emph{is} significant energy carried in the highly magnetized jet layers (either boundaries or internal regions), in the sense that the layers with $\beta_{pl}^{-1} >1$ may carry a significant amount of the jet total energy flux, from a few up to even tens of percent.

\subsection{Tangential Discontinuities}

In our analysis presented above, we have assumed that the jet magnetic and gaseous radial profiles are continuous. For such, we noted that sharp gradients in the comoving magnetic pressure $f(x)$ often generate $p(x)<0$. On the other hand, in the framework of the MHD approximation, tangential pressure discontinuities may be present, and those may elevate the overall jet $\sigma$ parameter without un-physical negative particle pressure.

The simplest parametrization of the $f(x)$ profile allowing for a sudden jump in the comoving magnetic pressure, is
\begin{equation}
\label{magneticField}
f(x) = A \, x^{-\varepsilon} \times H[x-x_0] \, ,
\end{equation}
with $\varepsilon>0$, where $H[x-x_0]$ is the Heaviside step function; note that the normalization $f(1)= 1$ implies $A\equiv 1$. The magnetohydrostatic equilibrium condition, along with the normalization $p(1) = 1$, then gives
\begin{eqnarray}
p(x) & = & 1 + \frac{2q}{\varepsilon} x_0^{-\varepsilon} - \frac{q (2-\varepsilon)}{\varepsilon} - \frac{2q}{\varepsilon} x_0^{-\varepsilon} \, H[x-x_0] + \nonumber \\
&& \frac{q (2-\varepsilon)}{\varepsilon} x^{-\varepsilon} \, H[x-x_0] \, ,
\end{eqnarray}
with $\varepsilon<2$. The discontinuity in the co-moving magnetic field profile at $x=x_0$ corresponds therefore to the discontinuity in the gaseous pressure. For example, $\varepsilon = 1$ corresponds to $f(x) = 0$ with $p(x) = 1 - q \,(1 - 2 x_0^{-1})$ for $x<x_0$, and $f(x) = x^{-1}$ with $p(x) = 1- q \, (1 - x^{-1})$ for $x>x_0$; at the same time, the total jet pressure $P_{\rm tot} \propto p(x) + q \, f(x)$ is continuous across the jet.

Let us moreover assume a uniform bulk velocity across the jet, $\Gamma(x) =$\,const, so that $h(x) \propto x$. This assumption is handy for two different reasons. Firstly, it significantly simplifies calculations and, secondly, a negligible velocity shear in fact always maximizes the overall jet $\sigma$ value (since the toroidal magnetic field pressure is typically larger in the outer layers of a jet). With such, we obtain
\begin{eqnarray}
\label{dis_sigma}
\sigma & = & \frac{q}{2} \,\, \frac{\int\limits_0^1\textrm{d}x \,h(x)\,f(x)}{\int\limits_0^1\textrm{d}x \,h(x)\,p\!(x)} =  \frac{q}{1+q} \,\, \frac{1-x_0^{2-\varepsilon}}{2-\varepsilon} \nonumber \\
&\equiv &  \Delta\!p \times  \frac{x_0^{\varepsilon}-x_0^{2}}{(1+q) \, (2-\varepsilon) } \, ,
\end{eqnarray}
where the gaseous pressure jump at $x=x_0$ is
\begin{equation}
\label{jump}
\Delta\!p \equiv  \lim_{x\to x_0-}p(x) - \lim_{x\to x_0+}p(x)= q \, x_0^{-\varepsilon}
\end{equation}

\begin{figure}
	\centering
	\includegraphics[width=0.75\columnwidth]{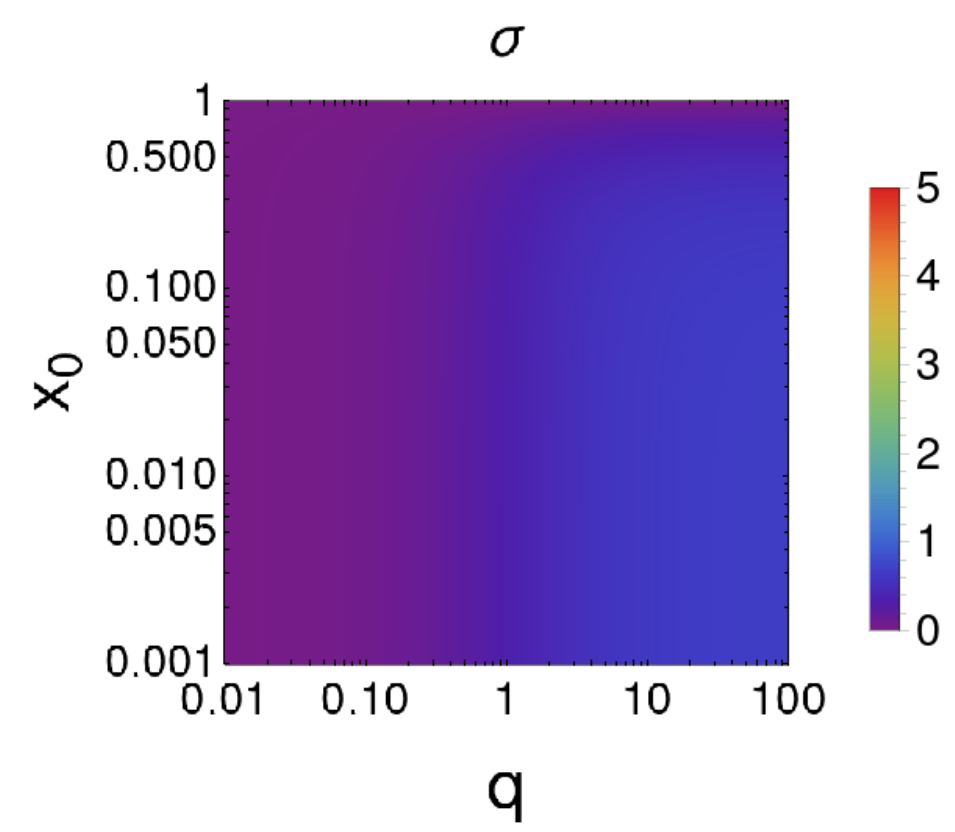}
	\includegraphics[width=0.75\columnwidth]{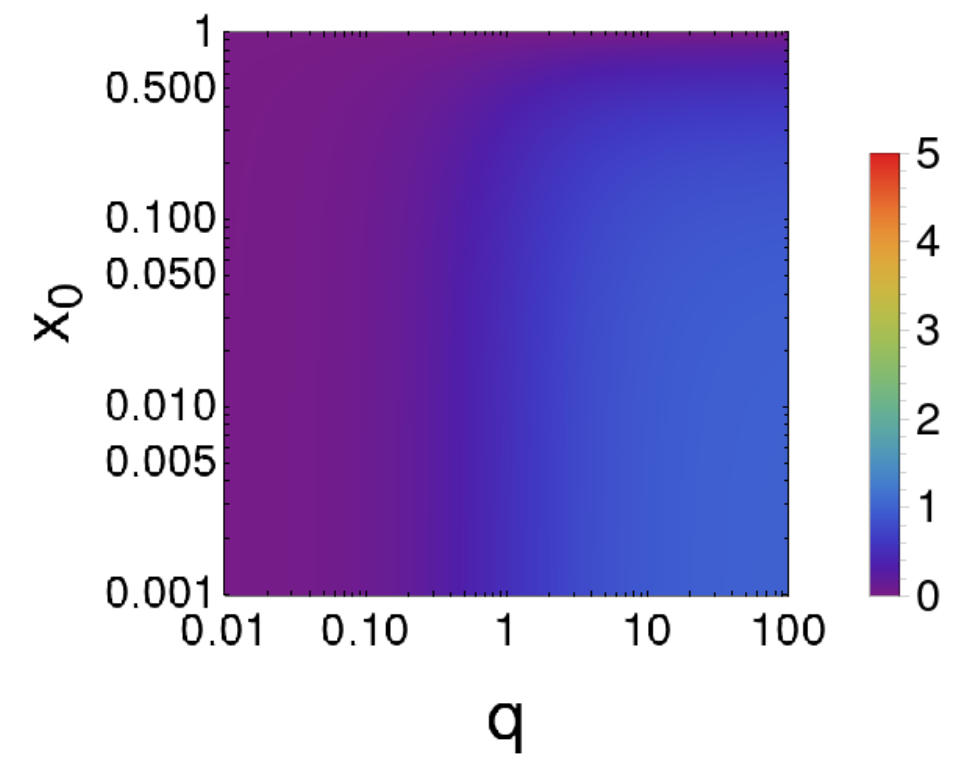}
	\includegraphics[width=0.75\columnwidth]{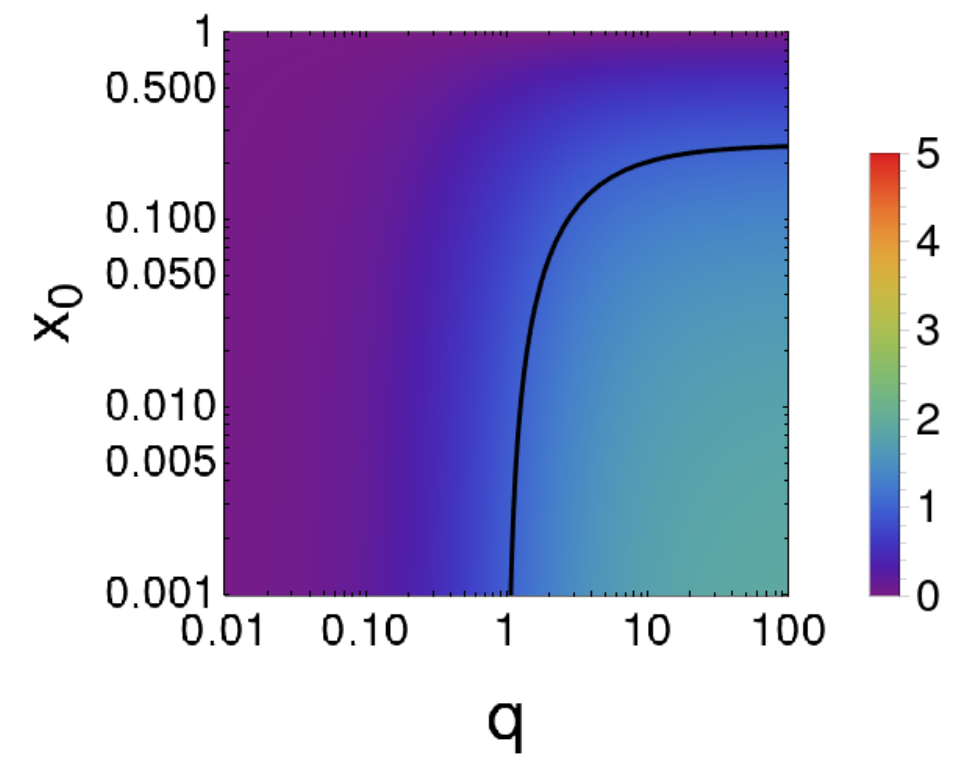}
	\includegraphics[width=0.75\columnwidth]{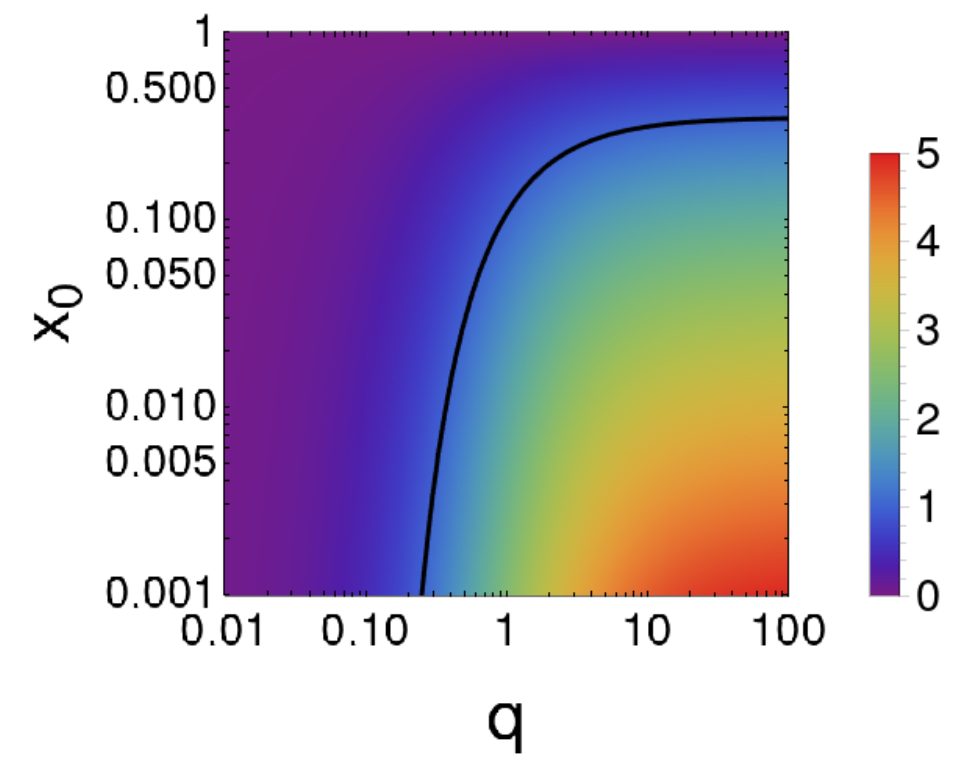}
	\caption{Density plots of $\sigma$ given in equation \ref{dis_sigma} as functions of $x_0$ and $q$, for the $\varepsilon$ values of 0.5, 1.0, 1.5, and 1.9 (from top to bottom, respectively). Black contours correspond to $\sigma=1$.}
	\label{sig}
\end{figure}

\begin{figure}
	\centering
	\includegraphics[width=0.8\columnwidth]{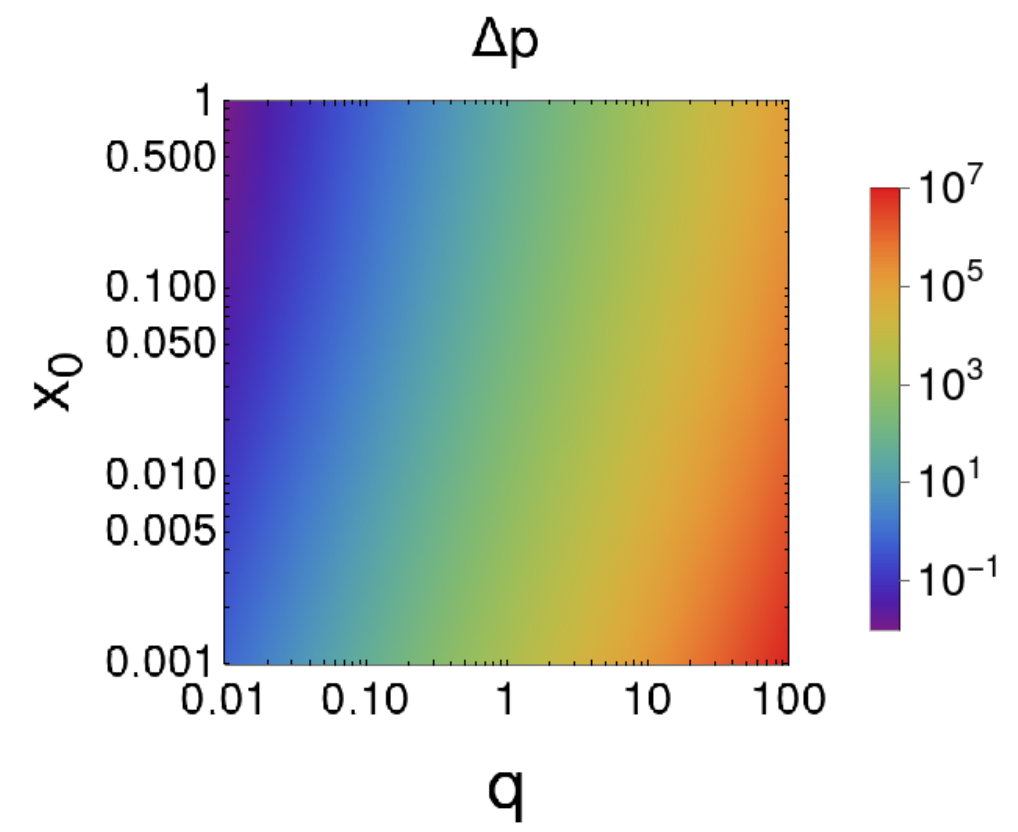}
	\includegraphics[width=0.8\columnwidth]{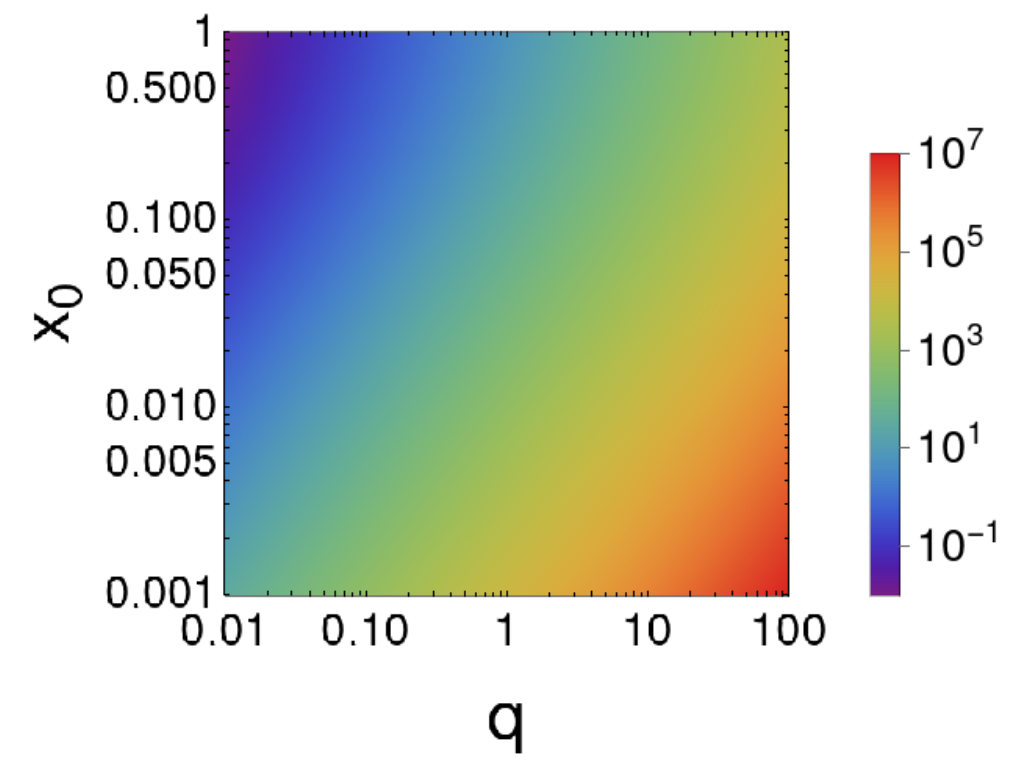}
	\includegraphics[width=0.8\columnwidth]{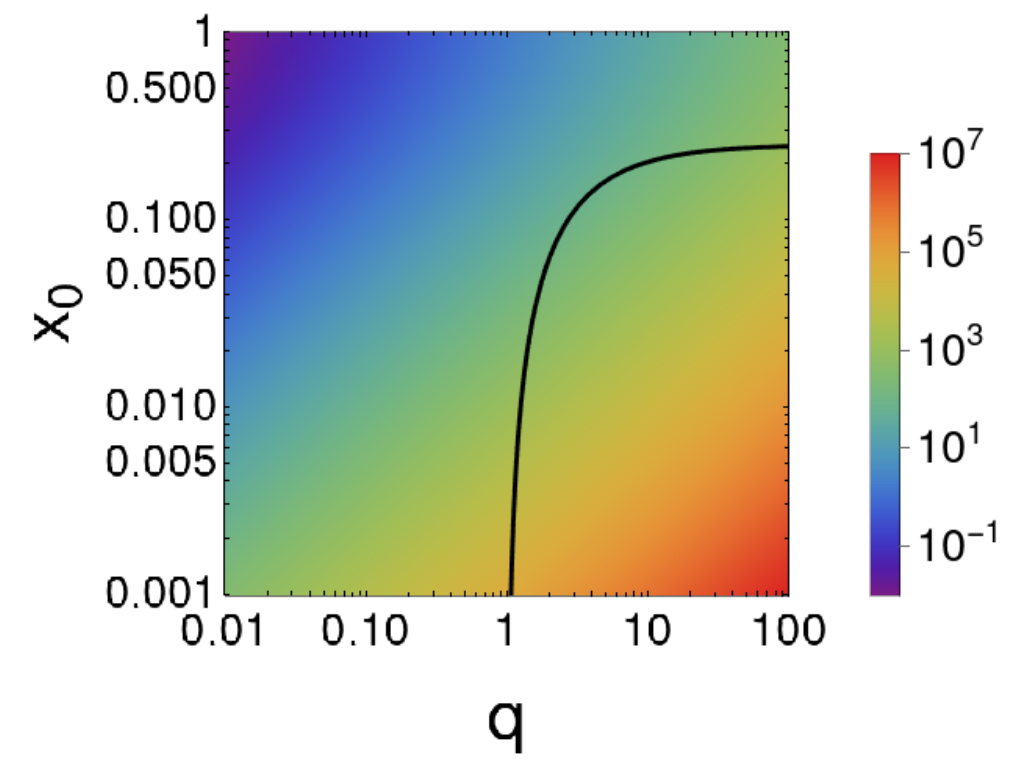}
	\includegraphics[width=0.8\columnwidth]{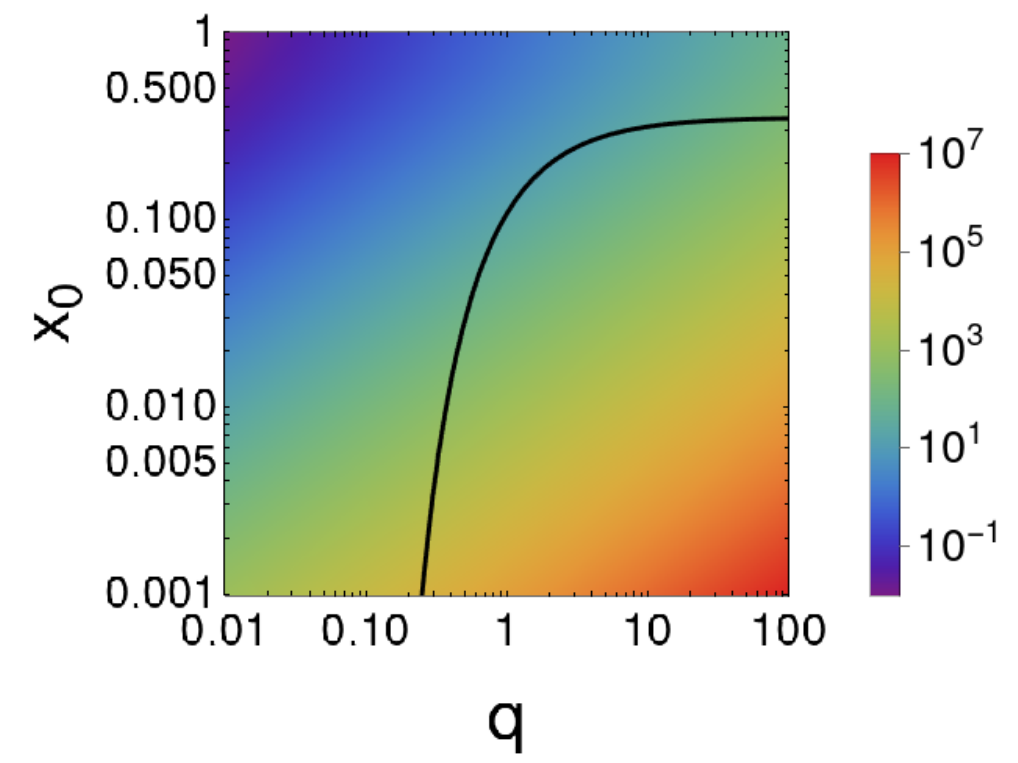}
	\caption{Density plots of  $\Delta\!p$ given in equation \ref{jump} as functions of $x_0$ and $q$, for the $\varepsilon$ values of 0.5, 1.0, 1.5, and 1.9 (from top to bottom, respectively). Black contours correspond to $\sigma=1$.}
	\label{dp}
\end{figure}

Clearly, in the above the $\sigma$ parameter is always smaller than unity for $\varepsilon \leq 1$. In the range $\varepsilon \in (1,2)$, for a given set of the $q$ and $x_0$ values, $\sigma$ increases with $\varepsilon$ approaching 2, and may occasionally exceed unity. Considering however that $\lim_{\varepsilon\to 2}\sigma=-\frac{q}{1+q}\ln x_0$, and also $\frac{q}{1+q} \sim 1$ for large $q>1$, we see that the jet magnetization increases only very slowly (logarithmically) with decreasing $x_0$. At the same time, the jump in the gaseous pressure $\Delta\!p$ increases very rapidly with decreasing $x_0$, $\varepsilon$ approaching 2, and finally with increasing $q$, as illustrated in Figures\,\ref{sig} and \ref{dp}. For example, with $\varepsilon=1.9$, $q=10$ and $x_0=0.01$, we have $\sigma\simeq 3.3$ and $\Delta\!p \sim 3 \times 10^3$; lowering $x_0$ by one order of magnitude results in only slightly higher $\sigma\simeq 4.5$, but dramatically larger $\Delta\!p \sim 3 \times 10^5$.

All in all, we conclude that, by allowing tangential discontinuities in the gaseous and magnetic pressure profiles, and at the same time diminishing radial velocity shear, one can elevate the jet magnetization parameter above unity. However, values of $\sigma \lesssim \mathcal{O}(10)$, require pressure jumps of several orders of magnitude, which could hardly be considered  realistic.\footnote{In fact, there is an intermediate class of profiles, for which the jet gaseous and magnetic pressures, albeit continuous, change with radius by orders of magnitude, while the jet magnetization parameter approaches/exceeds unity. In particular, consider a simple parametrization $f(x) = (1 + a) \, x^2 / (1 + a x^4)$ and a constant jet bulk Lorentz factor across the outflow. For such, $p(x) = 1 + q + (1+a)q/\sqrt{a}\, [\tan^{-1}(\sqrt{a}) -  \tan^{-1}(x^2 \sqrt{a})] - q f(x)$ and $\sigma=\frac{q}{1+q}(1+a)(4a)^{-1} \, \ln{(1+a)}$, so that with $q \to \infty$ and $a>50$ one can indeed obtain $\sigma \gtrsim 1$; for example, with $a=3,000$ and $q=10,000$, one has $\sigma \simeq 2$ and $p(0) \simeq 10^6$.}

\section{Discussion and Summary}
\label{sec:discussion}

In this paper we analyse the magnetization of a relativistic jet at large distances from the launching site, where it can be considered as fully formed, i.e. accelerated to terminal bulk velocity. In our simple model, we consider perfectly cylindrical jet geometry, purely toroidal configurations for the jet magnetic field, monotonic radial bulk velocity shear, and ultra-relativistic equation of state for the jet particles. We show analytically that, as long as the jet plasma is in magnetohydrostatic equilibrium, and the pressure radial profiles are continuous with the comoving magnetic pressure attaining its maximum at the jet boundary, the ratio of the electromagnetic to particle energy fluxes, both integrated over the jet cross-section area, has to be below unity, $\sigma < 1$. 

More complex cases, in particular those with global maxima of the magnetic pressure located well within the jet body, could only be explored numerically. For such, we found that sharp gradients in the comoving magnetic pressure often lead to unphysical solutions with negative particle pressure for some ranges of the jet radius. But when the particle pressure is positive everywhere, the condition $\sigma < 1$ tends to hold anyway. At the same time, however, for certain magnetic and bulk velocity profiles, magnetic pressure may still dominate over particle pressure for certain ranges of cylindrical radius within the jet. In other words, even though a current-carrying outflow as a whole is dominated by the kinetic energy flux of the jet particles, there may be extended domains carrying a significant fraction of the total jet energy, in which the comoving jet magnetic pressure dominates over the gaseous pressure, while the jet Poynting flux is comparable to the particle energy flux. This finding may be relevant in the context of jet particle acceleration processes, since energy dissipation in relativistic plasma proceeds rather differently depending on the plasma magnetization.

Recent Particle-in-Cell simulations find that magnetic reconnection may proceed very efficiently in the regime of high magnetization, $\beta_{pl} < 1$  and $\sigma >1$, where shocks are expected to be weakened by strong magnetic forces.  In both pair and electron-ion plasmas, these simulations exhibit efficient non-thermal particle acceleration, with maximum energies and power-law indices that depend on the particular values of the plasma magnetization parameters \citep[e.g.,][]{Sironi14,Guo15,Guo16,Werner16,Werner18,Petropoulou19}.

Our analysis therefore identifies an interesting possibility for astrophysical jets to be characterized by radial stratification with respect to both the plasma bulk velocity and the plasma magnetization. This includes not only the case of a particle-dominated jet spine surrounded by a magnetically-dominated boundary layer, but also the possibility of alternating-current jets consisting of layers with low and high values of $\sigma$ and $\beta_{pl}$. In such outflows, shocks and magnetic reconnection may dominate alternately, resulting in the formation of highly in-homogeneous distributions of radiating particles (with respect to maximum particle energies and particle spectral indices). When combined with relativistic beaming effects related to the radial velocity shear \citep[see, e.g.,][]{Komissarov90,Stawarz02,Aloy08}, this may lead to jets having vastly diverse appearances to an observer, depending on the jet viewing angle.

In the framework of the analyzed simple jet models with purely toroidal magnetic field, we have found that the jet magnetization parameter can be elevated up to relatively modest values $\sigma \lesssim \mathcal{O}(10)$ only in the case of extreme gradients or discontinuities in the gas pressure, \emph{and} a significantly suppressed velocity shear. Such cases would therefore correspond to a narrow, unmagnetized jet spine, surrounded by an extended, essentially force-free layer with a toroidal field, both characterized by comparable bulk Lorentz factors. However, in the absence of velocity shear, relativistic outflows with strong toroidal magnetic field (and no poloidal component) are known to be susceptible to current-driven kink instability (e.g., \citealt{Mizuno12,Nalewajko12,Marti16,Kim18,Das19,Sobacchi19} and references therein; for a recent review on the topic see also \citealt{Perucho19}).

Let us therefore comment in this context on the role played by an additional poloidal magnetic field, represented here for simplicity by a purely vertical component $B_z$. If this component is uniform across and along the jet, $\partial_{\phi} B_z  = \partial_z B_z = \partial_r B_z = 0$, the gaseous pressure profiles following from magnetohydrostatic equilibrium remain unchanged with respect to the ones we have calculated and discussed above, since no additional current component $\vec{J'} \propto \vec{\nabla'} \times \vec{B'}$ is associated with such an additional field, and so the magnetohydrostatic equilibrium $\vec{\nabla'} P \propto \vec{J'} \times \vec{B'}$ is not affected either. Moreover, the relative increase of the jet Poynting flux is then only in the $\phi$ direction, so it represents energy circulation around the jet and does not add to the original Poynting flux along the $z$ direction. Hence, there is no net increase in the $\sigma$ parameter. That is to say, a small amount of a uniform vertical field should stabilize the jet against the current-driven oscillations \citep[see][]{Mizuno12,Das19}, but should not affect our conclusions presented above regarding the jet magnetization.
	
On the other hand, in the presence of pronounced radial gradients in the vertical field, $\partial_r B_z \neq 0$ (but still  $\partial_{\phi} B_z  = \partial_z B_z = 0$), the situation may change. That is because, in such a case the particle energy flux $L_p \propto \int\!dr \, r \, \beta \, \Gamma^2 \, P $ should be affected as the gaseous pressure profile has to adjust in response to the additional azimuthal current component, $4 \pi \, J'_{\phi} = - c \,  \partial_{r'} B'_z$, following the altered magnetohydrostatic equilibrium, namely $\partial_r P \propto B_z \, \partial_r\!B_z  -  \frac{1}{2} r^{-2} \, \partial_r(r^2 B_{\phi}^2 \,\Gamma^{-2})$.
 
We note, however, that the configuration with the vertical field confined within a narrow jet spine, and therefore with strong radial gradients in $B_z$, is to be expected rather only for fully collimated and accelerated electromagnetic outflows, i.e. the ones for which the Poynting and particle energy fluxes are in a rough equipartition $\sigma \sim \mathcal{O}(1)$ anyway \citep[see][]{Beskin09,Lyubarsky09}. Moreover, as discussed in \citet{Mizuno12}, relativistic outflows consisting of a poloidal field concentrated toward the jet axis, and a dynamically relevant toroidal field in the outer layer, are highly kink-unstable, and as such subjected to efficient dissipation of the magnetic energy. Hence, we conclude that large values of the jet magnetization parameter $\sigma$ should not be realistically expected in such cases anyway.

\begin{acknowledgements}

D.{\L}.K. and {\L}.S. were supported by the Polish National Science Center grants 2016/22/E/ST9/00061 and DEC-2019/35/O/ST9/04054. M.C.B. acknowledges support from U.S. National Science Foundation grant AST 1903335. J.-M.M. and M.P. acknowledge support from the Spanish \emph{Ministerio de Ciencia} through grant PID2019-107427GB-C33, and from the \emph{Generalitat Valenciana} through grant PROMETEU/2019/071. J.-M.M. acknowledges additional support from the Spanish \emph{Ministerio de Econom\'{\i}a y Competitividad} through grant PGC2018-095984-Bl00. M.P. acknowledges additional support from the Spanish \emph{Ministerio de Ciencia} through grant PID2019-105510GB-C31.

\end{acknowledgements}

\bibliographystyle{aasjournal}

\begin{thebibliography}{99} 

\bibitem[Aloy \& Rezzolla(2006)]{Aloy06} Aloy, M.~A. \& Rezzolla, L.\ 2006, \apjl, 640, L115. doi:10.1086/503608

\bibitem[Aloy \& Mimica(2008)]{Aloy08} Aloy, M.~A. \& Mimica, P.\ 2008, \apj, 681, 84. doi:10.1086/588605

\bibitem[Appl \& Camenzind(1992)]{Appl92} Appl, S. \& Camenzind, M.\ 1992, \aap, 256, 354

\bibitem[Begelman et al.(1984)]{Begelman84} Begelman, M.~C., Blandford, R.~D., \& Rees, M.~J.\ 1984, Reviews of Modern Physics, 56, 255. doi:10.1103/RevModPhys.56.255

\bibitem[Begelman \& Cioffi(1989)]{Begelman89} Begelman, M.~C. \& Cioffi, D.~F.\ 1989, \apjl, 345, L21. doi:10.1086/185542

\bibitem[Beskin \& Nokhrina(2009)]{Beskin09} Beskin, V.~S. \& Nokhrina, E.~E.\ 2009, \mnras, 397, 1486. doi:10.1111/j.1365-2966.2009.14964.x

\bibitem[Blandford \& Znajek(1977)]{Blandford77} Blandford, R.~D. \& Znajek, R.~L.\ 1977, \mnras, 179, 433. doi:10.1093/mnras/179.3.433

\bibitem[Chatterjee et al.(2019)]{Chatterjee19} Chatterjee, K., Liska, M., Tchekhovskoy, A., et al.\ 2019, \mnras, 490, 2200. doi:10.1093/mnras/stz2626

\bibitem[Das \& Begelman(2019)]{Das19} Das, U. \& Begelman, M.~C.\ 2019, \mnras, 482, 2107. doi:10.1093/mnras/sty2675

\bibitem[Ghisellini et al.(2010)]{Ghisellini10} Ghisellini, G., Tavecchio, F., Foschini, L., et al.\ 2010, \mnras, 402, 497. doi:10.1111/j.1365-2966.2009.15898.x

\bibitem[Giannios \& Spruit(2006)]{Giannios06} Giannios, D. \& Spruit, H.~C.\ 2006, \aap, 450, 887. doi:10.1051/0004-6361:20054107

\bibitem[Guo et al.(2015)]{Guo15} Guo, F., Liu, Y.-H., Daughton, W., et al.\ 2015, \apj, 806, 167. doi:10.1088/0004-637X/806/2/167

\bibitem[Guo et al.(2016)]{Guo16} Guo, F., Li, H., Daughton, W., et al.\ 2016, Physics of Plasmas, 23, 055708. doi:10.1063/1.4948284

\bibitem[Kim et al.(2018)]{Kim18} Kim, J., Balsara, D.~S., Lyutikov, M., et al.\ 2018, \mnras, 474, 3954. doi:10.1093/mnras/stx3065

\bibitem[Kirk et al.(2000)]{Kirk00} Kirk, J.~G., Guthmann, A.~W., Gallant, Y.~A., et al.\ 2000, \apj, 542, 235. doi:10.1086/309533

\bibitem[Komissarov(1990)]{Komissarov90} Komissarov, S.~S.\ 1990, Soviet Astronomy Letters, 16, 284

\bibitem[Komissarov(1999)]{Komissarov99} Komissarov, S.~S.\ 1999, \mnras, 308, 1069. doi:10.1046/j.1365-8711.1999.02783.x

\bibitem[Komissarov \& Porth(2021)]{Komissarov21} Komissarov, S. \& Porth, O.\ 2021, \nar, 92, 101610. doi:10.1016/j.newar.2021.101610

\bibitem[Lisanti \& Blandford(2007)]{Lisanti07} Lisanti, M. \& Blandford, R.\ 2007, The First GLAST Symposium, 921, 343. doi:10.1063/1.2757343

\bibitem[Lyubarsky(2009)]{Lyubarsky09} Lyubarsky, Y.\ 2009, \apj, 698, 1570. doi:10.1088/0004-637X/698/2/1570

\bibitem[Lyubarsky(2010)]{Lyubarsky10} Lyubarsky, Y.~E.\ 2010, \mnras, 402, 353. doi:10.1111/j.1365-2966.2009.15877.x

\bibitem[Mart{\'\i} et al.(2016)]{Marti16} Mart{\'\i}, J.~M., Perucho, M., \& G{\'o}mez, J.~L.\ 2016, \apj, 831, 163. doi:10.3847/0004-637X/831/2/163

\bibitem[Meier(2012)]{Meier12} Meier, D.~L.\ 2012, Black Hole Astrophysics: The Engine Paradigm, by David L. Meier. ISBN: 978-3-642-01935-7. Springer, Verlag Berlin Heidelberg, 2012

\bibitem[Mizuno et al.(2008)]{Mizuno08} Mizuno, Y., Hardee, P., Hartmann, D.~H., et al.\ 2008, \apj, 672, 72. doi:10.1086/523625

\bibitem[Mizuno et al.(2012)]{Mizuno12} Mizuno, Y., Lyubarsky, Y., Nishikawa, K.-I., et al.\ 2012, \apj, 757, 16. doi:10.1088/0004-637X/757/1/16

\bibitem[Nalewajko \& Begelman(2012)]{Nalewajko12} Nalewajko, K. \& Begelman, M.~C.\ 2012, \mnras, 427, 2480. doi:10.1111/j.1365-2966.2012.22117.x

\bibitem[Perucho(2019)]{Perucho19} Perucho, M.\ 2019, Galaxies, 7, 70. doi:10.3390/galaxies7030070

\bibitem[Petropoulou et al.(2019)]{Petropoulou19} Petropoulou, M., Sironi, L., Spitkovsky, A., et al.\ 2019, \apj, 880, 37. doi:10.3847/1538-4357/ab287a

\bibitem[Rueda-Becerril et al.(2014)]{Rueda14} Rueda-Becerril, J.~M., Mimica, P., \& Aloy, M.~A.\ 2014, \mnras, 438, 1856. doi:10.1093/mnras/stt2335

\bibitem[Saito et al.(2015)]{Saito15} Saito, S., Stawarz, {\L}., Tanaka, Y.~T., et al.\ 2015, \apj, 809, 171. doi:10.1088/0004-637X/809/2/171

\bibitem[Sikora et al.(2005)]{Sikora05} Sikora, M., Begelman, M.~C., Madejski, G.~M., et al.\ 2005, \apj, 625, 72. doi:10.1086/429314

\bibitem[Sikora et al.(2009)]{Sikora09} Sikora, M., Stawarz, {\L}., Moderski, R., et al.\ 2009, \apj, 704, 38. doi:10.1088/0004-637X/704/1/38

\bibitem[Sironi \& Spitkovsky(2014)]{Sironi14} Sironi, L. \& Spitkovsky, A.\ 2014, \apjl, 783, L21. doi:10.1088/2041-8205/783/1/L21

\bibitem[Sobacchi \& Lyubarsky(2019)]{Sobacchi19} Sobacchi, E. \& Lyubarsky, Y.~E.\ 2019, \mnras, 484, 1192. doi:10.1093/mnras/stz044

\bibitem[Sobacchi et al.(2017)]{Sobacchi17} Sobacchi, E., Lyubarsky, Y.~E., \& Sormani, M.~C.\ 2017, \mnras, 468, 4635. doi:10.1093/mnras/stx807

\bibitem[Stawarz \& Ostrowski(2002)]{Stawarz02} Stawarz, {\L}. \& Ostrowski, M.\ 2002, \apj, 578, 763. doi:10.1086/342649

\bibitem[Werner et al.(2016)]{Werner16} Werner, G.~R., Uzdensky, D.~A., Cerutti, B., et al.\ 2016, \apjl, 816, L8. doi:10.3847/2041-8205/816/1/L8

\bibitem[Werner et al.(2018)]{Werner18} Werner, G.~R., Uzdensky, D.~A., Begelman, M.~C., et al.\ 2018, \mnras, 473, 4840. doi:10.1093/mnras/stx2530


\end{thebibliography}

\end{document}